\documentclass{aa}
\usepackage{graphicx}
\usepackage{txfonts}
\usepackage{natbib}
\bibpunct{(}{)}{;}{a}{}{,} 
\newcommand{\mincir}{\raise -2.truept\hbox{\rlap{\hbox{$\sim$}}\raise5.truept
\hbox{$<$}\ }}
\newcommand{\magcir}{\raise -2.truept\hbox{\rlap{\hbox{$\sim$}}\raise5.truept
\hbox{$>$}\ }}
\newcommand{\siml}{\raise -2.truept\hbox{\rlap{\hbox{$\sim$}}\raise5.truept
\hbox{$<$}\ }}
\newcommand{\simg}{\raise -2.truept\hbox{\rlap{\hbox{$\sim$}}\raise5.truept
\hbox{$>$}\ }}
\newcommand{\be}{\begin{equation}}
\newcommand{\ee}{\end{equation}}
\newcommand{\ba}{\begin{eqnarray}}
\newcommand{\ea}{\end{eqnarray}}
\newcommand {\kpc} {kpc $\;$}

\newcommand {\mpcc} {Mpc}

\newcommand {\h} {Mpc$\;$}

\newcommand {\ks} {km~s$^{-1} \;$}
\newcommand {\kss} {km~s$^{-1}$}

\newcommand {\mqua} {$\times 10^{14}\;M_{\odot} \;$}
\newcommand {\mquaa} {$\times 10^{14}\;M_{\odot}$}

\newcommand{\degree}{\ensuremath{\mathrm{^\circ}}}
\newcommand{\arcm}{\ensuremath{\mathrm{^\prime}\;}}
\newcommand{\arcs}{\ensuremath{\arcmm\hskip -0.1em\arcmm \;}}
\newcommand{\arcmm}{\ensuremath{\mathrm{^\prime}}}

\newcommand{\dotsec}{\,\rlap{\hbox{$\mathrm{^s}$}}{\hbox{$.$}}\,}

\begin{document}
\title{CLASH-VLT: Galaxy cluster MACS\,J0329-0211 and  its surroundings using galaxies
  as kinematic tracers}

%

\author{M. Girardi\inst{1,2},
W. Boschin\inst{3,4,5},
A. Mercurio\inst{6,7},
N. Nocerino\inst{1},
M. Nonino\inst{2}\thanks{We dedicate this paper to the memory of our friend and colleague Mario.},
P. Rosati\inst{8,9},
A. Biviano\inst{2,10},
R. Demarco\inst{11}
C. Grillo\inst{12,13},
B. Sartoris\inst{2,14}
P. Tozzi\inst{15}
E. Vanzella\inst{9}
}


\institute{Dipartimento di Fisica dell'Universit\`a degli Studi di Trieste -
Sezione di Astronomia, via Tiepolo 11, I-34143 Trieste, Italy \email{marisa.girardi@inaf.it}
\and INAF - Osservatorio Astronomico di Trieste, via Tiepolo 11,
I-34143 Trieste, Italy
\and Fundaci\'on Galileo Galilei - INAF (Telescopio Nazionale
  Galileo), Rambla Jos\'e Ana Fern\'andez Perez 7, E-38712 Bre\~na
  Baja (La Palma), Canary Islands, Spain
\and Instituto de Astrof\'{\i}sica de Canarias, C/V\'{\i}a L\'actea
s/n, E-38205 La Laguna (Tenerife), Canary Islands, Spain
\and Departamento de Astrof\'{\i}sica, Univ. de La Laguna, Av. del
Astrof\'{\i}sico Francisco S\'anchez s/n, E-38205 La Laguna
(Tenerife), Spain
\and Dipartimento di Fisica E.R. Caianiello, Universit\`a Degli Studi di Salerno, Via Giovanni Paolo II, I-84084 Fisciano (SA), Italy
\and
INAF - Osservatorio Astronomico di Capodimonte, via Moiariello 16, I-80131 Napoli, Italy
\and Dipartimento di Fisica e Scienze della Terra, Universit\`a  di Ferrara, via Saragat 1, 44122 Ferrara, Italy
\and INAF - Osservatorio di Astrofisica e Scienza dello Spazio di Bologna, via Gobetti 93/3, 40129 Bologna, Italy
\and IFPU - Institute for Fundamental Physics of the Universe, via Beirut 2, 34014 Trieste, Italy 
\and Institute of Astrophysics, Facultad de Ciencias Exactas, Universidad Andr\'es Bello, Sede Concepci\'on, Talcahuano, Chile 
\and Dipartimento di Fisica, Universit\`a degli Studi di Milano, via Celoria 16, 20133 Milano, Italy
\and INAF - IASF Milano, via A. Corti 12, 20133 Milano, Italy
\and University Observatory , Ludwig-Maximilians University Munich, Scheinerstrasse 1,81679 Munich, Germany
\and INAF - Osservatorio Astroﬁsico di Arcetri, Largo E. Fermi, 50125
Firenze, Italy
}

\date{Received  / Accepted }

\abstract {The study of substructure is an important step in
  determining how galaxy clusters form.}{We aim to gain new insights
  into the controversial dynamical status of MACS\,J0329-0211
  (MACS0329), a massive cluster at $z=0.4503\pm0.0003$, with a new
  analysis using a large sample of member galaxies as kinematic
  tracers.}  {Our analysis is based on extensive spectroscopic data
  for more than 1700 galaxies obtained with the VIMOS and MUSE
  spectrographs at the Very Large Telescope (VLT), in combination with
  $B$ and $R_{\rm C}$ Suprime-Cam photometry from the Subaru archive.
  According to our member selection procedure, we define a sample of
  430 MACS0329 galaxies within 6 Mpc, corresponding to $\sim$3 times
  the virial radius.}  {We estimate the global velocity dispersion,
  $\sigma_V=841_{-36}^{+26}$ \kss, and present the velocity dispersion
  profile. We estimate a mass $M_{\rm 200}=(9.2\pm 1.5)$ \mquaa, using
  227 galaxies within $R_{200} = (1.71\pm0.07)$ Mpc, for which
  $\sigma_{V,200} =1018_{-48}^{+40}$ \kss.  The spatial distribution
  of the red galaxies traces a SE-NW elongated structure, without
  signs of a velocity gradient.  This structure likely originates from
  the main phase of cluster assembly.  The distribution of the blue
  galaxies is less concentrated, more rounded and shows signs of
  substructure, all characteristics indicating a recent infall of
  groups from the field. We detect two loose clumps of blue galaxies
  in the south and southwest at a distance $\sim R_{200}$ from the
  cluster center. The strong spatial segregation among galaxy
  populations is not accompanied by kinematical difference. Thanks to
  our extensive catalog of spectroscopic redshift, we are able to
  study galaxy systems that are intervening along the line of
  sight. We can identify two foreground galaxy systems (GrG1 at $z\sim
  0.31$ and GrG2 at $z\sim 0.38$) and one background system (GrG3 at
  $z\sim 0.47$).  We point out that the second brightest galaxy
  projected onto the MACS0329 core is in fact the dominant galaxy of
  the foreground group GrG2.  MACS0329, GrG3 and two other systems
  detected using DESI DR9 photometric redshifts are close to each
  other, suggesting the presence of a large-scale structure.}
          {MACS0329 is close to a state of dynamical equilibrium,
            although surrounded by a very rich environment. We
            emphasize that the use of an extensive redshift survey is
            essential to avoid misinterpretation of structures
            projected along the line of sight.}


\titlerunning{Galaxy cluster MACS\,J0329-0211} 
\authorrunning{Girardi et al.} 
   \maketitle

\section{Introduction}
\label{intro}

In the context of cold dark matter $\Lambda$CDM cosmology, numerical simulations show
that structure formation takes place hierarchically and culminates in the
formation of galaxy clusters.  Galaxy clusters are preferentially
formed by anisotropic accretion along the large-scale-structure filaments and
to a considerable extent  by the accretion of galaxy groups, while
the merger of two or more massive entities is a rare case
(e.g., \citealt{berrier2009}; \citealt{mcgee2009}). Clusters are
multicomponent systems, and merger and accretion phenomena are detected
using multiwavelength observations and a variety of techniques focusing
on the different components, mainly dark matter, hot intracluster
medium (ICM) and galaxies (see \citealt{feretti2002};
\citealt{molnar2016}).

The use of multiband optical data, and in particular multiobject
spectroscopy, is a consolidated method to study the overall structure ans substructures of
galaxy clusters and cluster merging phenomena. This information
complements the X-ray information, since it is known that galaxies and
the ICM react on different timescales during a merger, as shown by
numerical simulations (e.g., \citealt{roettiger1997};
\citealt{springel2007}; \citealt{mcdonald2022}) and observational data
(e.g., the Bullett cluster; \citealt{markevitch2006}).  Nearly
collisionless dark matter and galaxies are expected to take longer to
reach the dynamical equilibrium state than the collisional ICM (e.g.,
\citealt{poole2006}).

Since the formation history varies greatly from cluster to cluster
(e.g., \citealt{berrier2009}), most observational studies focus on
individual systems.  Very large spectroscopic datasets are needed to
study clusters in projected phase space for different galaxy types and
to infer the formation history, as it has been shown for a few
clusters for which spectroscopic information is available for hundreds
of members (e.g., \citealt{czoske2002}; 
  \citealt{rines2003}; \citealt{demarco2010}; \citealt{owers2011};
\citealt{munari2014}; \citealt{girardi2015}; \citealt{sohn2019};
\citealt{mercurio2021}).

The object of this study, the galaxy cluster \object{MACS\,J0329-0211}
(hereafter MACS0329) at $z \sim 0.45$ owes its name to its inclusion
in the Massive Cluster Survey (MACS; \citealt{ebeling2001}). It is
also listed as an extended source in the ROSAT All-Sky Survey Bright
Source Catalog (RASS-BSC; \citealt{voges1999}). MACS0329 is a fairly
X-ray bright ($L_{\rm X,bol}=17\times 10^{44}$ erg s$^{-1}$;
\citealt{postman2012}), hot ($T_{\rm X}\sim7$--8 keV,
\citealt{cavagnolo2008}; \citealt{postman2012}) and massive cluster
with $M_{200}$\footnote{We refer to $R_{\Delta}$ as the radius of a
  sphere in which the average mass density is $\Delta$ times the
  critical density, at the redshift of the galaxy system; $M_{\Delta}$
  is the mass contained in $R_{\Delta}$.}  in the range of 7--13 \mqua
(\citealt{schmidt2007}; \citealt{donahue2014};
\citealt{umetsu2014}; \citealt{merten2015}; \citealt{umetsu2016}; 
  \citealt{umetsu2018}; \citealt{herbonnet2019}).

MACS0329 was also part of the project ``Cluster Lensing And Supernova
survey with Hubble'' (CLASH; \citealt{postman2012}). The CLASH X-ray
selection favors the inclusion of highly relaxed clusters, and indeed
the X-ray image of MACS0329 is unimodal (e.g.,
\citealt{yuan2020}\footnote{http://zmtt.bao.ac.cn/galaxy{\_}clusters/dyXimages/chandra.html};
see also our Fig.~\ref{figimage}). \citet{postman2012} refer to
MACS0329 as one of the eight over 25 CLASH clusters with a possible
substructure. Indeed, MACS0329 was classified as dynamically relaxed
by some authors at the time (\citealt{schmidt2007};
\citealt{allen2008}), but \citet{maughan2008} reported evidence for
substructure. More recent results are also controversial and suggest either a
relaxed (e.g., \citealt{mann2012}; \citealt{yuan2020}) or a
non-relaxed cluster (e.g., \citealt{sayers2013}; \citealt{mantz2015}).
MACS0329 shows a bright cool core with low entropy (e.g.,
\citealt{cavagnolo2008}; \citealt{sayers2013}), which is to be
expected for a relaxed galaxy cluster.  However, when analyzing the
central regions, \citet{ueda2020} find a spiral-like pattern in the
X-ray residual image that is consistent with a gas sloshing in the core,
which is likely related to an infalling subcluster.  Embedded in the
cool core is a luminous brightest cluster galaxy (hereafter BCG) with
very high UV emission and star formation rate (\citealt{donahue2015};
\citealt{fogarty2015}), blue color, and strong optical emission lines
(\citealt{green2016}). The BCG also shows radio emission, which is
likely due to an active galactic nucleus (AGN; \citealt{yu2018}).  In
addition, Giacintucci et al.  ({\citeyear{giacintucci2014};
  \citeyear{giacintucci2019}) have detected a radio mini-halo, a
  phenomenon that only occurs in cool-core clusters and is likely
  related to gas sloshing and AGN feedback (e.g.,
  \citealt{richard2020}). Finally, there is a second giant elliptical
  galaxy 40\arcs northwest of the BCG, which stimulated the idea that the
  MACS0329 could be undergoing a merger event (e.g.,
  \citealt{caminha2019}). Indeed, \citet{demaio2015} analyzed the
  diffuse intracluster light surrounding the BCG and concluded that
  the two galaxies have already interacted with each other.

The extensive bibliography shows the great interest in MACS0329 and
the controversial evidence for its substructure.  To date, however,
there is no study based on the kinematics of the member
galaxies. MACS0329 is part of the ESO Large Program ``Dark Matter Mass
Distributions of Hubble Treasury Clusters and the Foundations of
$\Lambda$CDM Structure Formation Models'', which aims to obtain
  the panoramic spectroscopic survey of the 13 CLASH clusters visible
  from ESO-Paranal, also known as CLASH-VLT\footnote{Website currently
    located at https://sites.google.com/site/vltclashpublic/.}
  campaign (\citealt{rosati2014}) based on data obtained at the Very Large Telescope (VLT).

This article is organized as follows. We describe our data and the
selection of member galaxies in Sects.~2 and 3.  Sections~4, 5 and 6
deal with the structure of MACS0329, that is the main properties, the
galaxy population and the substructure. Other galaxy systems
projected onto the field of MACS0329 or in its vicinity are analyzed
in Sect.~7. Section~8 is devoted to the interpretation and discussion
of our results.  Our summary and conclusions can be found in Sect.~9.

In this work we use $H_0=70$ km s$^{-1}$ Mpc$^{-1}$ in a flat
cosmology with $\Omega_0=0.3$ and $\Omega_{\Lambda}=0.7$. In the
assumed cosmology, $1\arcm \sim 346$ \kpc at the
redshift of MACS0329.  All magnitudes are given in the AB system.  The
velocities we derive for the galaxies are line-of-sight velocities
determined from the redshift, $V=cz$. Unless otherwise stated, we report the
errors with a confidence level (c.l.) of 68\%.

\section{Data and galaxy catalog}
\label{data}

\begin{figure*}
\centering 
\includegraphics[width=17cm]{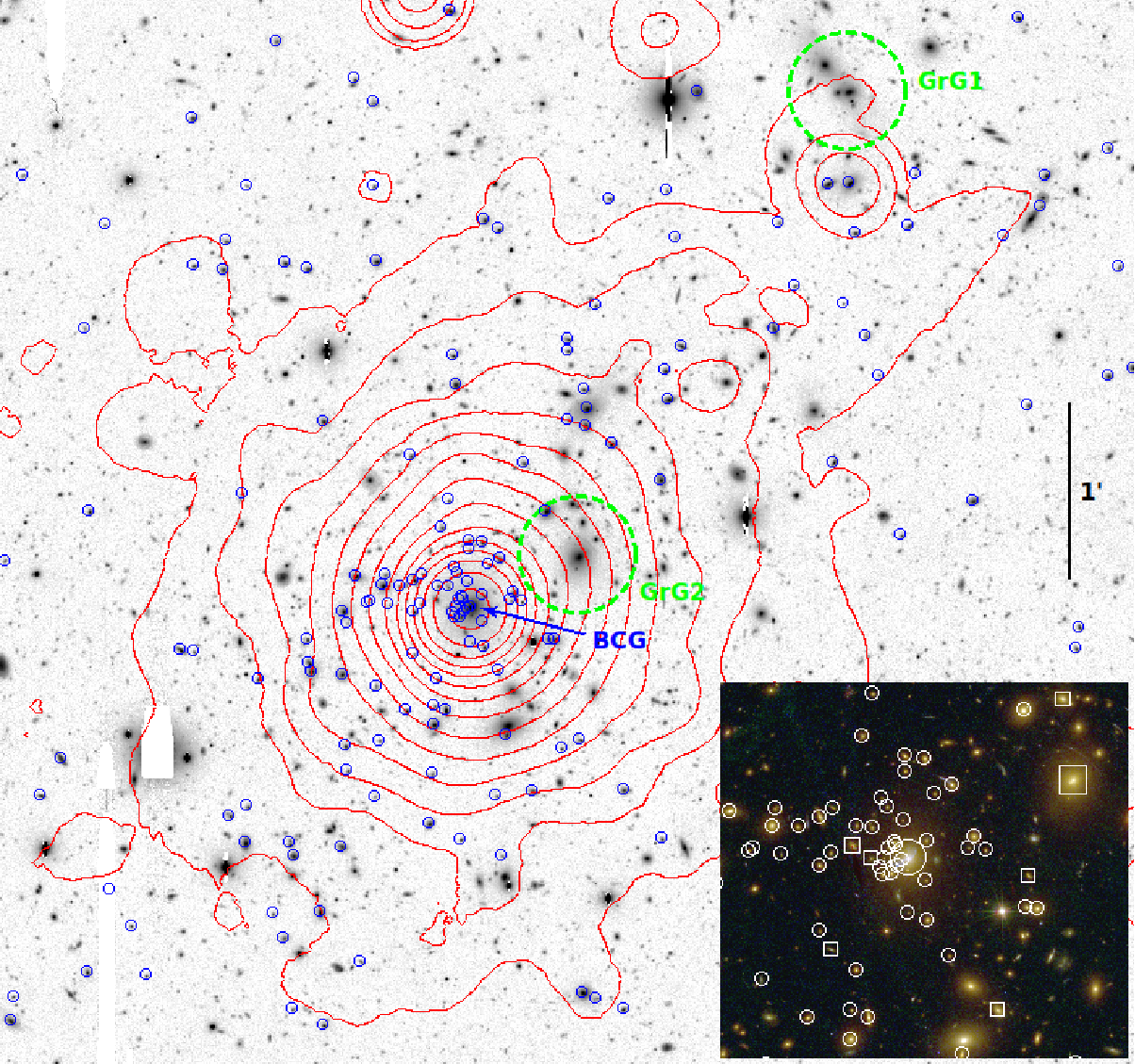}
\caption{Suprime-Cam $R_{\rm C}$-band image (north top and east left)
  of the MACS0329 field with superimposed isocontour levels of the
  Chandra X-ray emission (energy range: 0.5-7 keV). Only the main
  cluster region is shown. Small blue circles mark cluster members
  (see Sect.~\ref{memb}). The blue label indicates the brightest
  cluster galaxy. Green labels and large circles indicate the two
  foreground groups (GrG1 at $z=0.314$ and GrG2 at $z=0.385$)
  projected onto the field of the cluster (see Sect.~\ref{field}). The
  inset shows a color composite image of the central region
  ($1.5\times1.5$ arcmin$^2$) from the CLASH survey
  (\citealt{postman2012}), which combines HST/ACS and WFC3
  passbands. Member galaxies of MAC0329 and GrG2 are indicated by circles and
  squares, respectively (large symbols indicate the two brightest
  members of MACS0329 and GrG2).}
\label{figimage}
\end{figure*}

The spectroscopic catalog of MACS0329 here considered consists of 1712
galaxies with a redshift between 0 and 1.  We exploit an extensive
spectroscopic campaign of the MACS0329 field with the VIMOS
spectrograph (VIsible MultiObject Spectrograph), as part of the CLASH-VLT large program
(\citealt{rosati2014}), augmented with MUSE integral field
spectroscopy in the cluster core. For the full catalog we refer to
Rosati et al. (in prep.).  VLT-VIMOS data were reduced
with the VIPGI package (\citealt{scodeggio2005}), details can be found
in \citet{mercurio2021}.  Each redshift was assigned a quality flag.
In this work, only VIMOS redshifts with a quality flag QF$=3$ (secure)
and QF$=2$ were considered, the latter refer to redshifts with a
reliability $\ga 80\%$. The reliability of the quality flags is
  assessed on the basis of duplicate observations with at least one secure
  measurement, as described in \citet{balestra2016}.  Details about
VLT-MUSE data can be found in \citet{caminha2019}.  When both VIMOS
and MUSE redshifts are available, we only consider the MUSE
measurement.  A MUSE redshift measurement is available for the
brightest cluster galaxy.  In this study, all MUSE and VIMOS galaxies
with a redshift between 0 and 1 are considered, giving a total of 74
and 1637 redshifts, respectively. The VIMOS data provide us with an
efficient coverage of an important cluster volume sampling MACS0329
out to a radius $R\ \simg 3 R_{200}$. The MUSE data allow us to sample
the central $\siml 2\arcm$ region, which could not be adequately
sampled by standard multi-slit spectroscopy.

We also obtained two long-slit spectra for the very bright galaxy
$\sim 40$\arcs northwest of the BCG [at
  R.A.=$03^{\mathrm{h}}29^{\mathrm{m}}39\dotsec14$, Dec.=$-02\degree
  11\arcmm 29.2\arcs$ (J2000.0)].  In Sect.~\ref{field}, we show that
this galaxy, the second brightest galaxy projected onto the cluster
core, is in fact the dominant galaxy of a foreground group (GrG2; see Sect.~\ref{field}).  This
galaxy was observed with the Italian Telescopio {\it Galileo} (TNG) with
a total exposure time of 3600s in 2018 (see Fig.\ref{figspectrum}).

\begin{figure}
\centering 
\resizebox{\hsize}{!}{\includegraphics[angle=90]{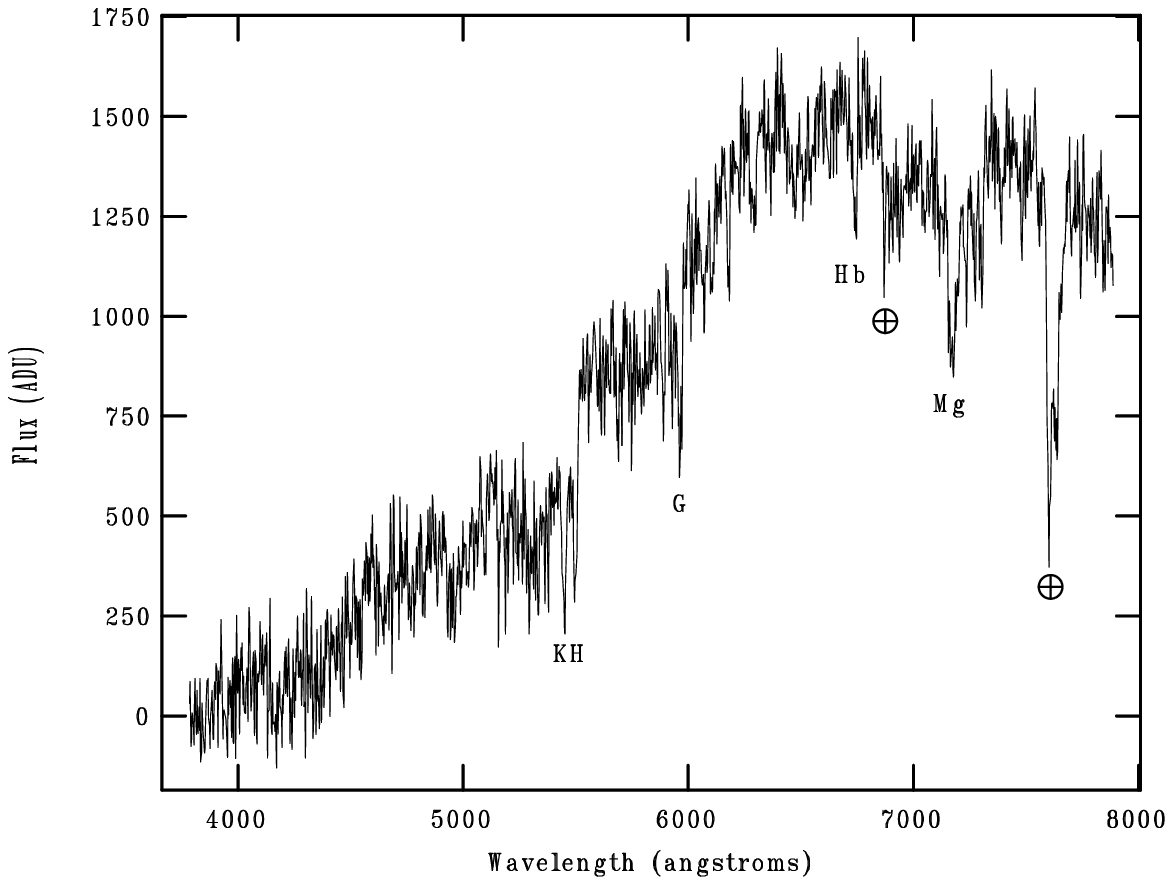}}
\caption{TNG spectrum of the brightest galaxy of the foreground GrG2 group
  ($z= 0.385$), which is projected $\sim 40$\arcs to the northwest of
  the center of MACS0329 (see Fig.~\ref{figimage}). The x-axis
    shows the rest-frame wavelength. The main spectral features are marked
    by their symbols. The two telluric features at $\sim 6900$~\textup{~\AA} and
    7600~\textup{~\AA} are also shown (marked by the earth symbol).}
  \label{figspectrum}
\end{figure}

For the MACS0329 field, we used photometric information from
Suprime-Cam data.  The images were retrieved from the CLASH
page\footnote{https://archive.stsci.edu/prepds/clash/} available at
the Mikulski Archive for Space Telescopes (MAST).  A full description
of the reduction of the Suprime-Cam images can be found in the data
section of the CLASH
website\footnote{https://archive.stsci.edu/missions/hlsp/clash/macs0329/data/subaru/}.
Briefly, the images were reduced by one of us using the techniques
described in \citet{nonino2009} and \citet{medezinski2013}.  The total
area covered by the images is $34\arcmm \times 27$\arcmm.  The
zeropoints are in the AB system.  Specifically, we retrieved the image
in the $R_{\rm C}$-band with an exposure time of $2400$~sec and a
depth of $26.5$~mag, and the image in the $B$-band with an exposure of
$2940$~sec and a depth of $27.2$~mag. The photometric catalogs were
produced using the software SExtractor (\citealt{bertin1996}). We
extracted independent catalogs in each band that were then matched
across the wavebands using STILTS (Starlink Tables Infrastructure Library Tool Set, \citealt{taylor2006}). Among the photometric
quantities, we measured aperture magnitudes (MAG{\_}APER) in
circular apertures with diameters of 5.0 arcsec and the Kron magnitude
(MAG{\_}AUTO), which is estimated through an adaptive elliptical
aperture (\citealt{kron1980}).  Unless otherwise stated, the values listed are
dereddened magnitudes, that is we have applied the correction for
Galactic extinction.

We were able to assign magnitudes to 1693 out of 1712 galaxies in the
redshift catalog. This  catalog is electronically
published in Table~\ref{tabcatesempio}, available at the Strasbourg astronomical Data Center (CDS).
VIMOS was used with both the
low-resolution blue grism (LRb) and the medium-resolution grism (MR)
with typical uncertainties of $cz=150$ and $75$ \kss, respectively
(\citealt{biviano2013}; \citealt{balestra2016}). Typical uncertainties
of MUSE redshifts are $cz=40$ \ks (\citealt{inami2017}). For the
TNG redshift, the uncertainty is $cz=100$ \kss.

\begin{table}[ht]
    \caption[]{Spectroscopic redshift catalog.}
\label{tabcatesempio}
           $$
           \begin{array}{r c c c c c c}
            \hline
            \noalign{\smallskip}
            \hline
            \noalign{\smallskip}

\mathrm{ID} & \alpha(\mathrm{J2000})&\delta(\mathrm{J2000})& z & R_{\mathrm{C}}& \mathrm{Mem}& \mathrm{Orig} \\
            & \mathrm{degree}&\mathrm{degree}&  &
\mathrm{mag}& \\

         \hline
         \noalign{\smallskip}
  1 &  52.352504 &   -2.288108  & 0.4540 & 23.84 &1&  VLR\\
  2 &  52.281052 &   -2.197914  & 0.3388 & 20.67 &0&  VLR\\
  3 &  52.254059 &   -2.066159  & 0.2797 & 22.36 &0&  VLR\\
  \noalign{\smallskip}
  \hline
           \end{array}
$$\tablefoot{This table is available at the CDS. A portion
             is shown here for guidance regarding its form and
             content. Col.~1: running ID for galaxies in the
             sample; Cols.~2 and 3: R.A. and Dec. (J2000);
             Col.~4: spectroscopic redshift, $z$; Col.~5: Kron $R_{\rm
               C}$-band magnitude;
             Col~6: MACS0329 member assignment
             (1: yes, 0: no).
             Col.~7: origin of spectroscopic data (VLR:
             VIMOS-VLT with LRb grism, VMR: VIMOS-VLT with MR grism, VMU:
             MUSE-VLT, TNG: TNG).
           }
\end{table}

\section{Member selection}
\label{memb}

To select cluster members from the 1712 galaxies with redshifts, we
applied the two-step method known as ``peak +gap'' (P+G), which has already
been used for CLASH-VLT clusters (e.g., \citealt{girardi2015}).  In the
first step, we analyzed the redshift sample using the 1D
adaptive-kernel method of Pisani et al. (\citeyear{pisani1993},
hereafter DEDICA).  Using this method, MACS0329 is identified as a peak at
$z=0.4504$ populated by 533 galaxies (in the range $0.4333\leq
z \leq 0.4664$, see Fig.~\ref{fighistonew}).

The redshift space around MACS0329 is particularly rich in structures,
and Table~\ref{tabdedica1d} lists all peaks with a relative
density with respect to the densest peak, $\rho>$0.25. These densest
peaks are associated with other galaxy systems and will be investigated in
Sect.~\ref{field}.

\begin{figure}
\centering
\resizebox{\hsize}{!}{\includegraphics{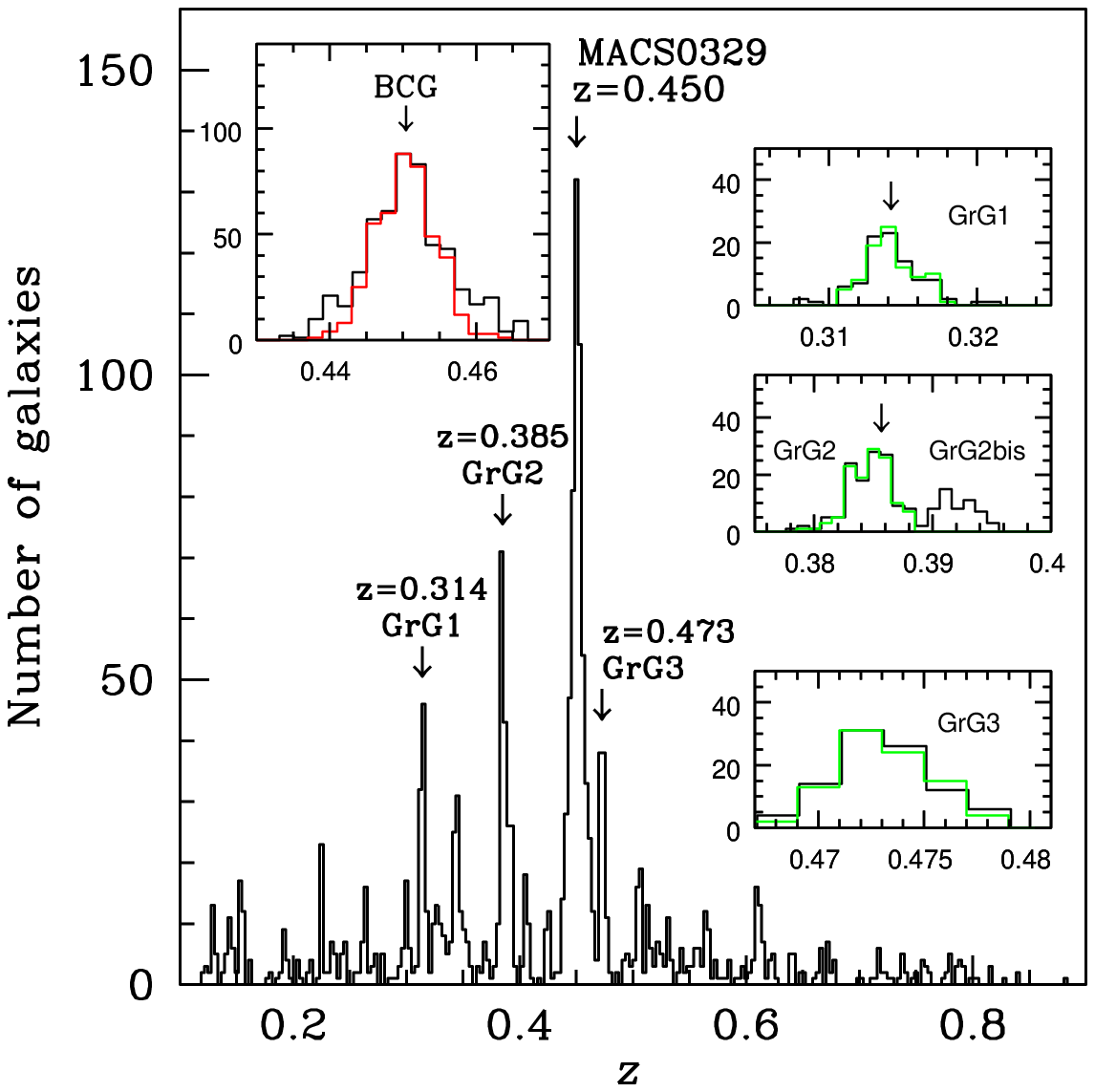}}
\caption
{Distribution of galaxy redshifts. The histogram refers to the
  galaxies with the spectroscopic redshift in the region of MACS0329
  with $0.1<z<0.9$. Labels indicate redshifts of the relevant peaks
  and the names of the corresponding galaxy systems, that is MACS0329
  and other groups (see Table~\ref{tabdedica1d}).  The redshift
  distribution of the 533 galaxies assigned to the MACS0329 peak can
  be seen in the inset plot at the top left (black line
    histogram). The histogram with the red line refers to the 430
    galaxies that are members. The BCG redshift is also shown. For
    each of the other groups, the respective right-hand inset panel
    shows the redshift distribution of the galaxies assigned to the
    peak and the galaxies selected as members (black and green line
    histograms, respectively, see Sect.~\ref{field}). For GrG1 and
    GrG2, the arrows indicate the redshifts of the luminous galaxies
    that we have selected as the centers of the groups.}  
\label{fighistonew}
\end{figure}

\begin{table}
        \caption[]{Peaks in the redshift distribution.}
         \label{tabdedica1d}
            $$
         \begin{array}{l c r c c l }
            \hline
            \noalign{\smallskip}
            \hline
            \noalign{\smallskip}
\mathrm{Peak} & z-\mathrm{range}&N_{\rm g} & z &\rho& \mathrm{Notes}\\
         \hline
         \noalign{\smallskip}
 \mathrm{no.~1}  &0.3077,0.3210&   95&  0.3140& 0.33&\mathrm{2D bim.,GrG1}\\
 \mathrm{no.~2}  &0.3780,0.3954&  182&  0.3850& 0.50&\mathrm{1D bim.,GrG2}\\
 \mathrm{main}   &0.4333,0.4664&  533&  0.4504& 1.00&\mathrm{MACS0329}\\
 \mathrm{no.~3}  &0.4673,0.4790&   93&  0.4728& 0.28&\mathrm{GrG3}\\
              \noalign{\smallskip}
            \hline
         \end{array}
         $$ \tablefoot{ Col.~1: peak ID; Col.~2: number of assigned
           member galaxies, $N_{\rm g}$; Col.~3: redshift of the
           density peak; Col.~4: relative density with respect to the
           highest peak, $\rho$; Col.~5: notes on the
           structures we detect (see Sect.~\ref{field} for bimodal
           peaks.}
\end{table}


\begin{figure}
\centering
\resizebox{\hsize}{!}{\includegraphics{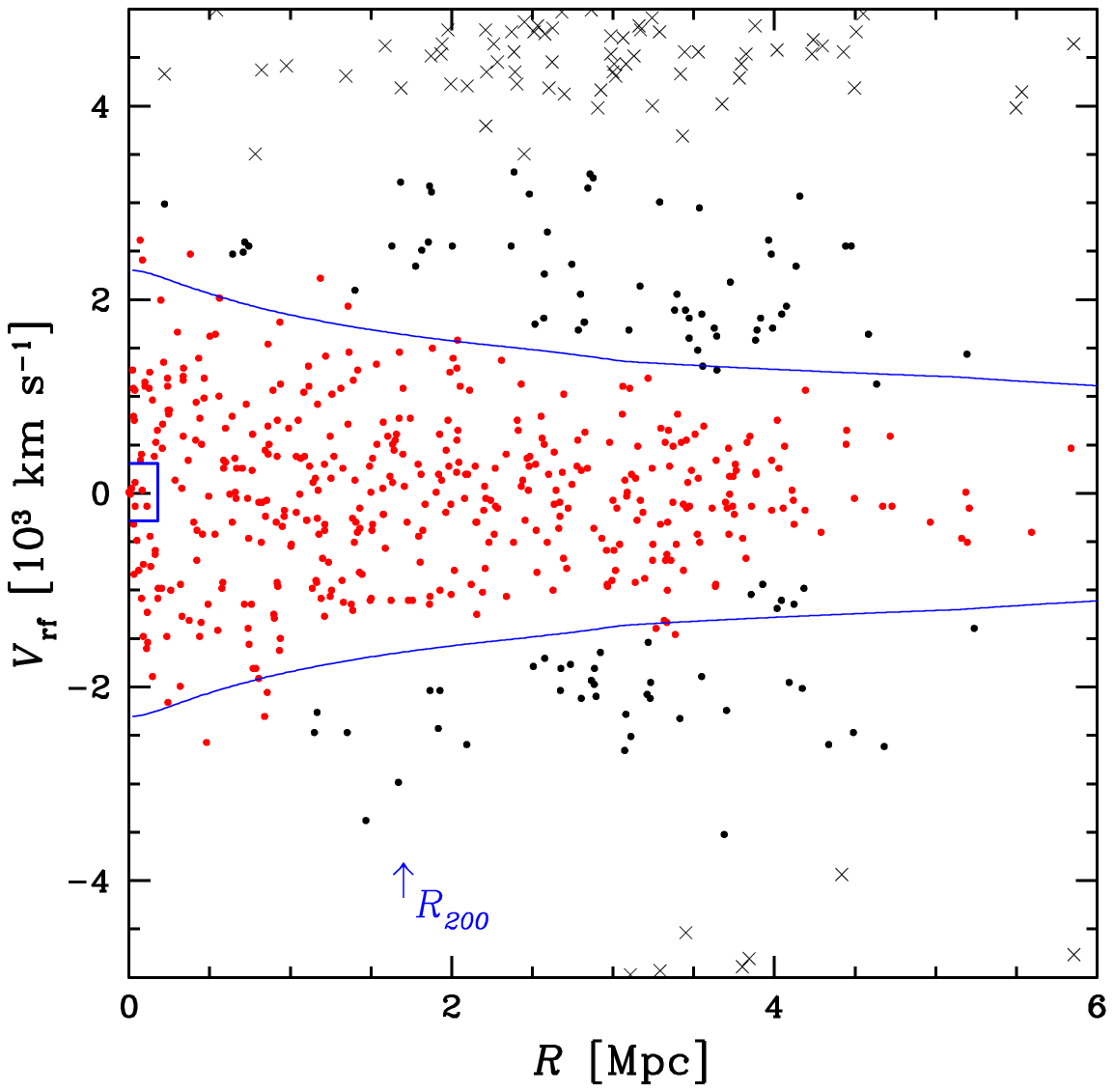}}
\caption
    {Projected phase space of galaxies in the MACS0329 field with the
      rest-frame velocity $V_{\rm rf}=(V-<V>)/(1+z)$ versus the
      projected clustercentric distance $R$. Dots show the 533
      galaxies selected in the velocity peak of the cluster according
      to the 1D-DEDICA method. Red dots indicate the 430 galaxies that
      are considered true cluster members according to the second step
      of the member selection.  The BCG position is indicated by the
      box at $R=0$. The blue curves contain the region where
      $V_{\rm rf}$ is smaller than the escape velocity (see text).  }
\label{figvd}
\end{figure}

\begin{table*}
        \caption{Global properties of MACS0329.}
         \label{tabv}
            $$
         \begin{array}{c c c c c c c c}
            \hline
            \noalign{\smallskip}
            \hline
            \noalign{\smallskip}

N_{\rm g} &{\rm R.A.(J2000)},\,{\rm Dec.(J2000)}&z&\sigma_{V}&N_{R200}&\sigma_{V,200}&R_{200}&M_{200}\\
  &\mathrm{h:m:s,\degree:\arcmm:\arcs}& &\mathrm{km\ s^{-1}}&&\mathrm{km\ s^{-1}}&{\rm Mpc}&10^{14}M_{\odot}\\
         \hline
         \noalign{\smallskip}
430&03\ 29\ 41.57-02\ 11\ 46.4&0.4503\pm0.0003&841_{-36}^{+26}&227 &1018_{-48}^{+40} &1.71\pm0.07&9.2\pm 1.5\\
              \noalign{\smallskip}
            \hline
         \end{array}
         $$ \tablefoot{Col.~1: the number of member galaxies of MACS0329,
           $N_{\rm g}$; Col.~2: the cluster center, for which we assume
the position of the BCG; Cols.~3 and 4: the mean redshift and
velocity dispersion computed using all $N_{\rm g}$ galaxies, $z$ and $\sigma_V$, with respective errors;
Col.~5: the number of member galaxies within $R_{200}$, $N_{R200}$; Col.~6:
the velocity dispersion computed using galaxies within $R_{200}$, $\sigma_{V,200}$, with its error;  Cols.~7 and 8: the $R_{200}$ radius and the mass there contained, $M_{200}$, with respective errors. 
$R_{200}$ and related quantities are determined using a recursive procedure
(see Sect.~\ref{prop}).}
         \end{table*}


In the second step, we combined the spatial and velocity information
which works in projected phase space.  In fact, galaxies in
  clusters appear within regions delimited by caustics with a
  characteristic trumpet shape (\citealt{kaiser1987};
    \citealt{regos1989}) and outside the caustics the density of
    galaxies drops substantially (see Fig.~\ref{figvd} for
    MACS0329). Galaxies outside the caustics are background or
    foreground galaxies (e.g., \citealt{denhartog1996};
    \citealt{diaferio1997}; \citealt{geller1999}). We applied the
  ``shifting gapper'' method (Fadda et al. \citeyear{fadda1996};
  Girardi et al. \citeyear{girardi1996}).  Of the galaxies that lie
  within an annulus around the center of the system, the method
  excludes those that are too far away in velocity from the main body
  of galaxies, that is further than a certain velocity distance called
  the velocity gap. The position of the annulus is shifted with
  increasing distance from the center of the cluster.  The procedure
  is repeated until the number of cluster members converges to a
  stable value. We defined the center of the cluster as the
    position in R.A. and Dec. of the BCG.  The BCG lies at the center
  of the main galaxy peak in the 2D spatial galaxy distribution (see
  Sect.~\ref{2D} and Fig.~\ref{figk2z}). Following \citet{fadda1996},
  we chose an annulus size of 0.6 Mpc or more to include at least 15
  galaxies.  For the velocity gap they suggested a value of $1000$
  \kss, but smaller values are more appropriate for deeply sampled
  clusters as in the case of CLASH-VLT (e.g., 800 \ks in
  \citealt{girardi2015} and $500$ \ks in \citealt{balestra2016}).

The projected phase space around MACS0329 is particularly rich in
galaxies (see Fig.~\ref{figvd}).  When applying different values for
the velocity gap, the difference in membership is small in the
$R_{200}$ cluster region, but significant in the outer cluster
regions.  To better understand the best value for the velocity gap, we
analyzed the velocity distribution of 236 galaxies that survived the
velocity gap of 800 \ks in the region outside 2 \mpcc. We used the
1D-Kaye's Mixture Model test (1D-KMM test; Ashman et
al. \citeyear{ashman1994}), which fits a user-defined number of
Gaussian distributions to a dataset and evaluates the improvement of
this fit compared to a single Gaussian distribution. A three-group
partition (18, 180 and 38 galaxies) is significantly better at the
$>99.9\%$ c.l. and proves that the cluster is indeed surrounded by
galaxies with low and high velocities.  We adopted a velocity gap of
500 \kss, leading to a membership in the outer cluster regions that is
very similar to the KMM result, that is differing for six of 180
galaxies.  Of the 533 galaxies in the cluster velocity peak, 103
galaxies are discarded by this procedure.  In summary, the sample of
fiducial cluster members consists of 430 galaxies (hereafter called
the TOT sample) of which 421 have complete photometric information.

Figure~\ref{figvd} also shows the escape velocity curves calculated
according to den Hartog \& Katgert (\citeyear{denhartog1996}). In
particular, we assumed a NFW mass density profile 
  (Navarro, Frenk and White - \citealt{navarro1997}) where the concentration parameter is given
by \citet{umetsu2018} and the mass value is estimated in the following
section.  This is an a posteriori verification of our membership
procedure.

\section{Global properties of the cluster}
\label{prop}

We calculated the mean redshift of the 430 cluster galaxies of the TOT sample
sample as $\left<z\right>=0.4503\pm0.0003$, which corresponds to a
mean line-of-sight velocity of $\left<{V}\right>=135\,011\pm$40 \kss,
using the biweight estimator (ROSTAT software for robust statistics by
\citealt{beers1990}).  We estimated the velocity dispersion,
$\sigma_{V}$, by applying the cosmological correction and the
standard correction for velocity errors (\citealt{danese1980}), thus obtaining  $\sigma_{V}=841_{-36}^{+26}$ \kss, where  errors have been estimated by a bootstrap method.

To derive the mass $M_{200}$, we used the theoretically expected
relation between $M_{200}$ and the velocity dispersion, which was also
verified on simulated clusters (Eq.~1 of \citealt{munari2013}) We
adopted a recursive approach. To obtain an initial estimate of the
radius $R_{200}$ and $M_{200}$, we applied the relation of Munari et
al. to the global value of $\sigma_{V}$.  We considered the galaxies
within this first estimate of $R_{200}$ to recalculate the velocity
dispersion.  The procedure is repeated until a stable result is
obtained. We estimate $\sigma_{V,200}=1018_{-48}^{+40}$ \kss for 227
galaxies within $R_{200}=(1.71\pm 0.07)$ Mpc, and $M_{200}=(9.2\pm
1.5)$ \mquaa.  The uncertainty for $R_{200}$ is calculated using the
error propagation for $\sigma_{V}$ ($R_{200}\propto \sigma_{V}$) and
the uncertainty for $M_{200}$ is estimated using a similar error
propagation ($M_{200}\propto \sigma_{V}^3$) and an additional
uncertainty of $10\%$ to account for the scatter around the relation
of \citet{munari2013}.
In the following, we refer to the 227 galaxies
within $R_{200}$ as the R200 sample. Of these 227 galaxies, 218 have
the full magnitude information.  The main properties of the cluster
are listed in Table~\ref{tabv}.

We have determined the cluster mass profile, $M(r)$, in two more
  sophisticated ways, namely using the MAMPOSSt method of Mamon et
  al. (\citeyear{mamon2013}) for Modeling Anisotropy and Mass Profiles of Observed Spherical Systems, and the Caustic method of Diaferio \&
  Geller (\citeyear{diaferio1997}). MAMPOSSt solves the Jeans equation
  for dynamical equilibrium by performing a maximum likelihood fit of 
  models  of cluster mass profile, $M(r)$, and velocity anisotropy profile,
  $\beta(r)$, to the distribution of cluster members in the
  projected phase-space. We chose the NFW model for $M(r)$ and the
  Tiret model (\citealt{tiret2007}) for $\beta(r)$.
We performed the fit to the number density profile of cluster members
externally to MAMPOSSt, by considering a NFW profile in projection
(\citealt{bartelmann1996}). We found a best-fit scale radius for the
number density profile $r_{\nu}=1.1_{-0.2}^{+0.3}$ Mpc.
We then determined the marginal distributions of the four free
parameters of $M(r)$ (the virial and scale radii, $R_{200}$ and $r_s$)
and $\beta(r)$ (the central velocity anisotropy $\beta_0$ and the
asymptotic value at large radii $\beta_{\infty}$) using the Markov
chain Monte Carlo (MCMC) technique. The resulting $M(r)$ best-fit
parameters and 68\% marginalized errors are
$R_{200}=1.57_{-0.09}^{+0.10}$ Mpc and $r_s=0.96_{-0.05}^{+0.08}$
Mpc, and those of $\beta(r)$ are $\beta_0=0.4_{-0.9}^{+0.4}$ and
$\beta_{\infty}=0.7_{-0.4}^{+0.5}$, indicating slightly radial orbits
for the cluster members.

The Caustic method is based on the identification of density
discontinuities in the projected phase-space and does not require the
assumption of a model for $M(r)$. The amplitude of the caustic in
velocity space as a function of radius can be converted to a mass
profile, modulo a factor ${\cal F}_{\beta}$. We have assumed that
${\cal F}_{\beta}=0.5$, as recommended by \citet{diaferio1997} and
Rines et al. (\citeyear{rines2003}, \citeyear{rines2013}). We have
calculated the uncertainties for the caustic $M(r)$ according to
\citet{diaferio1999} and \citet{serra2011}. We found $R_{200}=1.64 \pm
0.24$ Mpc.

The MAMPOSSt and Caustic mass profiles are shown in
Fig.~\ref{figmass}. In the same figure we also mark the determination
of ($R_{200}, M_{200}$) from the scaling relation with velocity
dispersion as computed above and other determinations of ($R_{200},
M_{200}$) from gravitational lensing and X-ray, as presented in the
literature. Most of the values agree within their error bars and the
value of $M_{200}$ resulting from the scaling relation with velocity
dispersion is intermediate among the different determinations. We
adopt this value, which we list in Table~\ref{tabv}, for the rest 
of the article.

\begin{figure}
\centering 
\resizebox{\hsize}{!}{\includegraphics{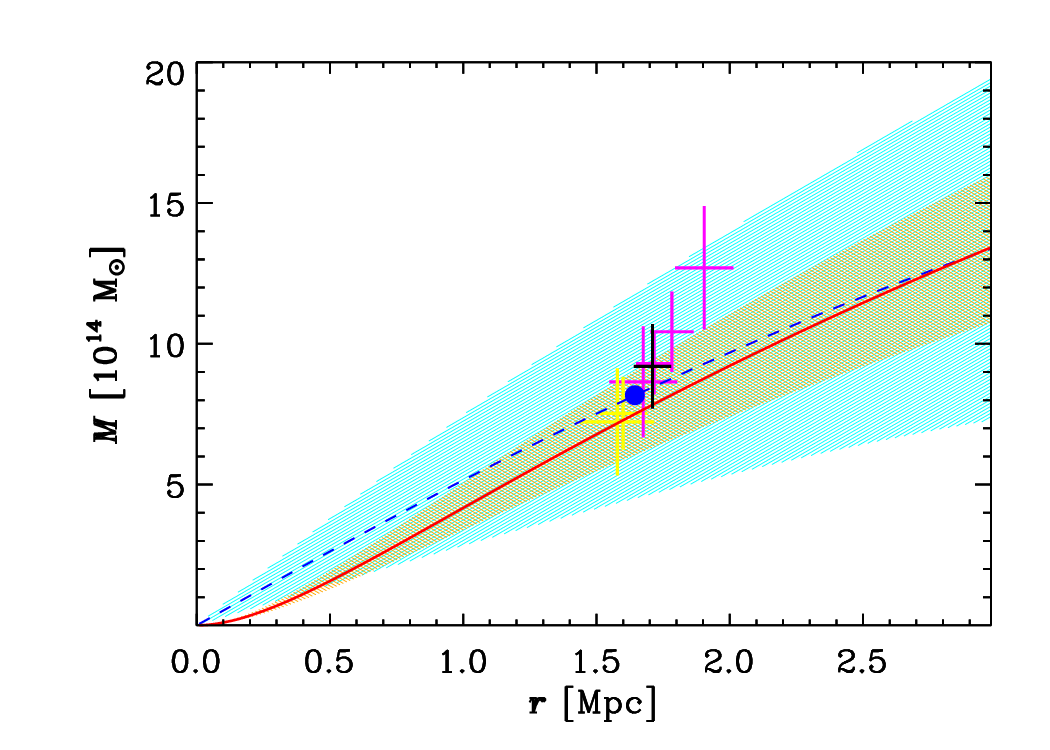}}
\caption
{Cluster mass profile $M(r)$. MAMPOSSt solution: solid red line
    (best fit) and orange shading (68\% confidence level). Caustic
    solution: dashed blue line (best fit) and cyan shading (68\%
    confidence level). The dots represent the positions of $(R_{200},
    M_{200})$ in the MAMPOSSt (red) and Caustic (blue) $M(r)$. The
    crosses stand for other determinations of $(R_{200}, M_{200})$
    with the respective one $\sigma$ error bars. The black cross is
    our determination based on the scaling relation with the cluster
    velocity dispersion (the value also given in Table
    1). The two yellow crosses are determinations based on the
    hydrostatic equilibrium of the intracluster X-ray emitting gas
    and the the four magenta crosses are determinations based on
    gravitational lensing effect. The references to these values are, from
    left to right in the figure: \citet{schmidt2007};
    \citet{donahue2014}; \citet{umetsu2016}); \citet{herbonnet2019};
    \citet{merten2015}; \citet{umetsu2018}.}
\label{figmass}
\end{figure}

Figure~\ref{figvdprof} (middle and bottom panels) shows the mean
velocity profile and the velocity dispersion profile for all member
galaxies. As for the integral profiles, the first value on the left
was calculated from the sample of the five galaxies closest to the
cluster center, and the following values were calculated by adding the
galaxies at larger radii one by one. The last values correspond to the
global estimates of $\left<V\right>$ and $\sigma_{V}$ which were
calculated for all cluster members.

\begin{figure}
\centering 
\includegraphics[width=15.5cm]{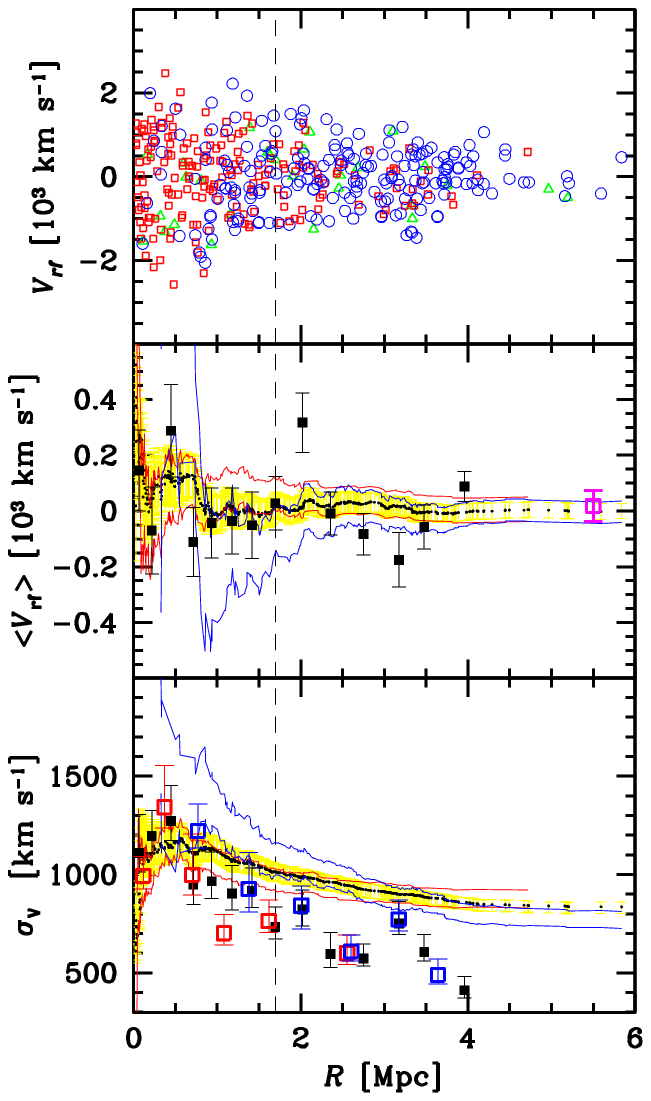}
\caption
    { Projected phase space and profiles of mean velocity and velocity
      dispersion for the member galaxies of MACS0329.  {\em Top
        panel:} projected phase space. Red, green and blue symbols
      show red, intermediate and blue galaxies as classified in
      Sect.~\ref{red} (418 galaxies; see Fig.~5).  {\em Middle panel:}
      mean rest-frame velocity profile for the 430 member
      galaxies (black full squares with error bars). Each value was
      computed using 30 galaxies.  The small black dots show the
      integral profile of the mean velocity and converge by definition
      towards the global value $\left<{V_{\rm
            rf}}\right>=0$. The error ranges are shown by yellow
        lines.  The magenta point indicates the rest-frame BCG
        velocity with its 2$\sigma$ uncertainty. It is located at a
        large radius for an easy comparison with the global value of
        the mean velocity.  For the integral profiles calculated for
        blue and red galaxies, only the 1$\sigma$ errors are shown
        (red and blue bands respectively).  {\em Bottom panel:} as
        above, but for the velocity dispersion profile. The integral
        profile for all galaxies (small black dots) converges towards
        the global value of $\sigma_{V}$. Red and blue open squares
        indicate differential profiles for red and blue galaxies.  The
        vertical dashed lines in the three panels indicate the
        $R_{200}$ radius.}
\label{figvdprof}
\end{figure}

\section{Red and blue galaxy populations}
\label{red}

Figure~\ref{figcm} shows the position of the 421 cluster members with
available magnitudes in the $B-R_{\rm C}$ vs. $R_{\rm C}$
color-magnitude diagram that we used to separate red and blue
galaxies.  The diagram shows the presence of the two usual overdense
regions, that along the red sequence and that of the blue cloud.  
  We emphasize that our magnitudes are not k-corrected magnitudes.  We
  have verified that our subsequent classification below (based on the
  color $B-R_{\rm C}$) is very similar to the one we would obtain using
  the color $B-Z$, which roughly corresponds to a rest frame NUV-$r$.
  \citet{wyder2007} gave the NUV-$r$ vs. $r$ color-magnitude diagram
  to clearly separate galaxies into red/early and blue/late
  populations.  We prefer to use $B-R$ for the comparison with the other two
  CLASH-VLT clusters, MACS~J1206.2-0847 at similar redshift
  (\citealt{girardi2015}) and Abell~S1063 at $z\sim 0.35$
  (\citealt{mercurio2021}), for which the color type classification shows
  good agreement with the respective spectroscopic type classification
  (see their Figs.~2 and 12).  The brightest cluster galaxies are usually
close to the red sequence or slightly bluer if there is relevant star
formation activity. In MACS0329 the BCG is characterized by a rather
blue color, $(B-R_{\rm C})\sim({\rm NUV}-B)_{\rm rf}\sim 1.4$,
which can be explained by a strong star formation activity probably
related to the presence of a cool core (see \citealt{fogarty2015}
  for the UV emission and other references in Sect.~\ref{intro}).

To fit the color-magnitude relation (CMR), we considered only the 269
galaxies redder than $B-R_{\rm C}$=1.3 and applied a $2\sigma$
rejection procedure.  The fitted relation is $B-R_{\rm
  C}=3.554(\pm0.123)-0.067(\pm 0.006)\times R_{\rm C}$, based on 151
galaxies. The fitted CMR is similar to that obtained by
\citet{girardi2015} for the cluster MACS~J1206.2-0847, as expected since
the two clusters have a similar redshift (see  their
Fig.~2). The color $(B-R_{\rm C})_{\rm diff}$ is defined as the
difference between the color of an observed galaxy and the
corresponding CMR color at the magnitude of the galaxy.

\begin{figure}
\centering
\resizebox{\hsize}{!}{\includegraphics{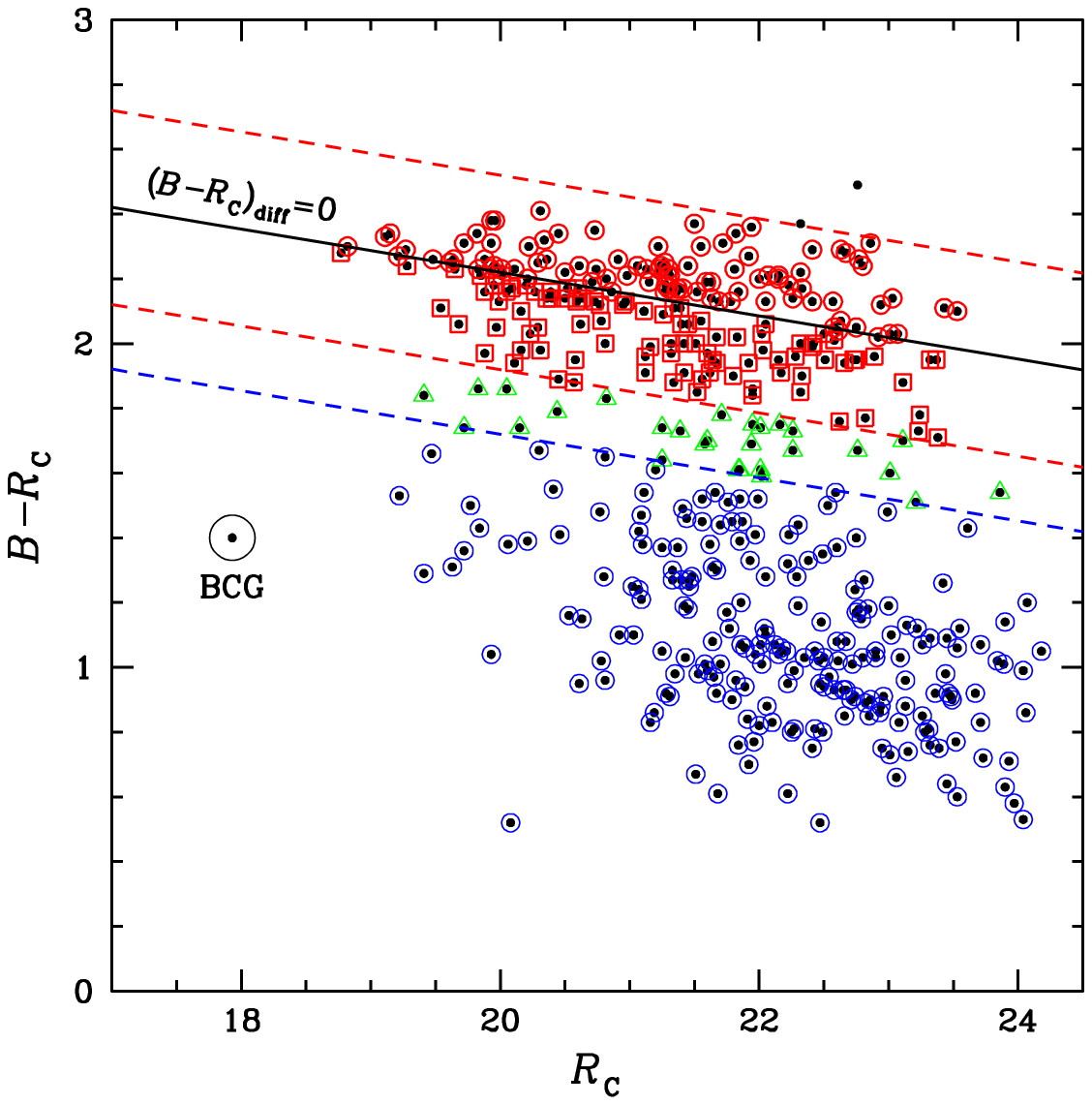}}
\caption {Suprime-Cam aperture-color vs. Kron-like magnitude diagram
  $B-R_{\rm C}$ vs. $R_{\rm C}$ for the 421 spectroscopic cluster members with
  photometric information.  The black solid line is the
  color-magnitude relation, $(B-R_{\rm C})_{\rm diff}=0$. The blue
  dashed line shows the value $(B-R_{\rm C})_{\rm diff} = -0.5$
  which we use to separate green and blue galaxies (highlighted by
  green triangles and blue circles).  The two red dashed lines show
  the locus of the red sequence galaxies, $|(B-R_{\rm C})_{\rm diff}|
  \le 0.3$, with the galaxies of the RedU and RedD samples marked
  with red circles and squares. A large black circle
  highlights the BCG.}
\label{figcm}
\end{figure}

Following \citet{girardi2015}, we classified the galaxies as
follows.  The Red sample consists of the galaxies with $|(B-R_{\rm
  C})_{\rm diff}| \le 0.3$ (189 galaxies), and among them, the galaxies
of the RedU and RedD samples are those with positive or negative
$(B-R_{\rm C})_{\rm diff}$ (94 and 95 galaxies, respectively).  We
then defined the Blue sample as the one containing 201 galaxies with
$(B-R_{\rm C})_{\rm diff} \le -0.5$.  Galaxies with intermediate
colors form the Green sample (28 galaxies with $-0.5<(B-R_{\rm
  C})_{\rm diff}<-0.3$).  The BCG is neither included in the Blue
sample nor in other color class sample.  In total, we classified 418
galaxies, of which 214 lie within $R_{200}$.

Figure~\ref{fig2d} shows the spatial distribution of all member
galaxies and in particular the fact that red galaxies are
spatially more clustered than blue galaxies.  To make a quantitative
comparison among different galaxy populations, we applied the
Kruskall-Wallis test to the clustercentric distances $R$ (KW test;
e.g., \citealt{lederman1984}). This test is a non-parametric procedure
that can be used to test whether three or more samples come from the same parent
distribution. In this case, the hypothesis is rejected with a c.l. of more
than $99.99\%$.  Figure~\ref{figKWr} shows the significant difference
among the galaxies in the Red, Green and Blue samples, with the
galaxies in the first groups being more clustered.  We also compared
the distributions of the clustercentric distances of the RedD and RedU
samples using the 1D Kolmogorov-Smirnov test (hereafter 1DKS test;  \citealt{kolmogorov1933}; \citealt{smirnov1948}; see also 
\citealt{lederman1984}) and found no significant difference.  We
also found no significant difference when comparing the positions in
RA and Dec of RedU and RedD galaxies according to the 2D
Kolmogorov-Smirnov test (hereafter 2DKS test; \citealt{fasano1987}).

\begin{figure}
\centering 
\resizebox{\hsize}{!}{\includegraphics{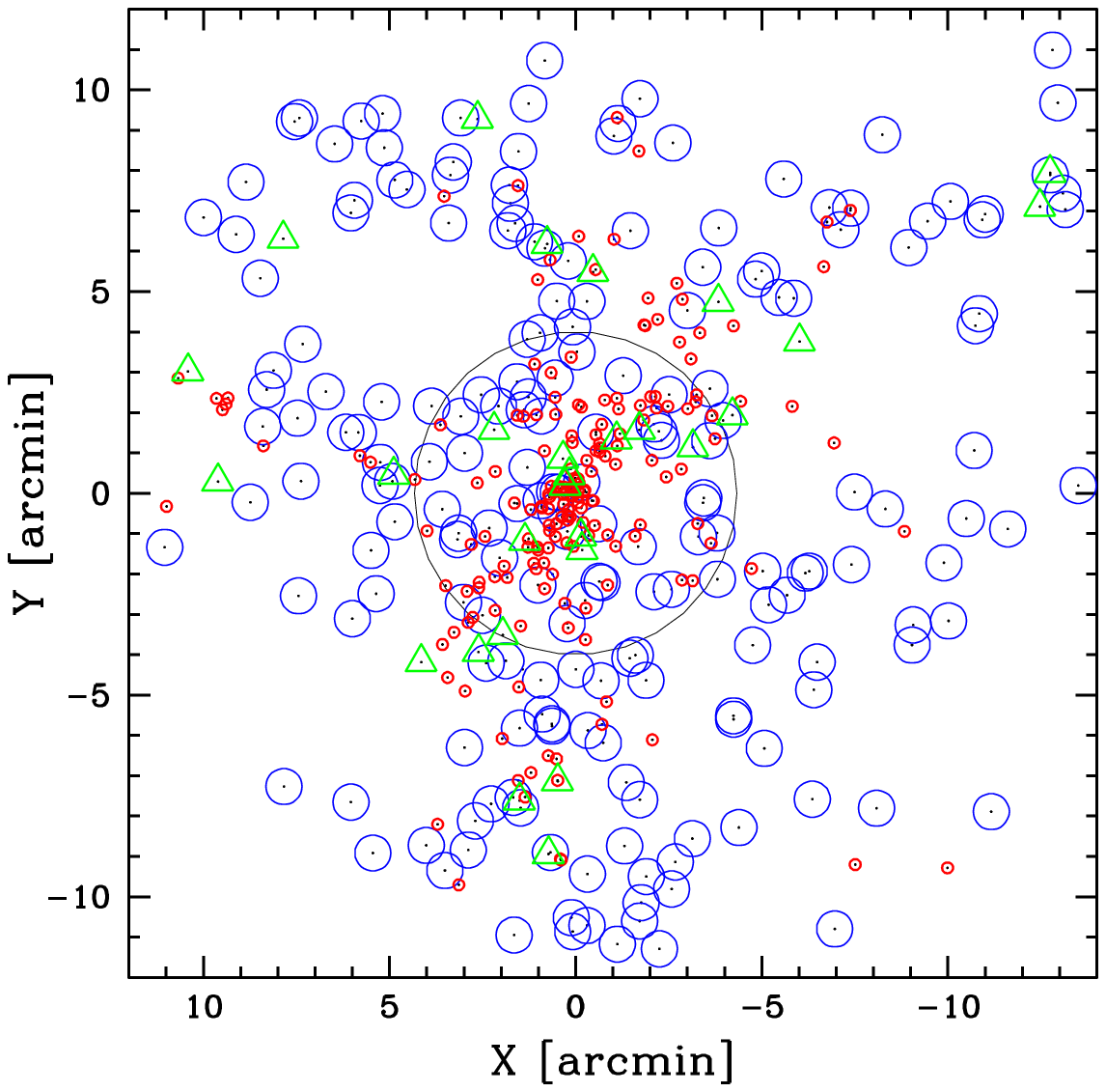}}
\caption
    {Spatial distribution of the 430 cluster members with emphasis on
      the spatial segregation among galaxies of different colors.
      Each one of the 418 classified galaxies is identified by a symbol:
      Red sample (small red circles); Green sample (green triangles);
      Blue sample (blue circles).  The large circle is centered on the
      BCG and encloses the $R_{200}$ region.}
\label{fig2d}
\end{figure}

\begin{figure}
\centering
\resizebox{\hsize}{!}{\includegraphics{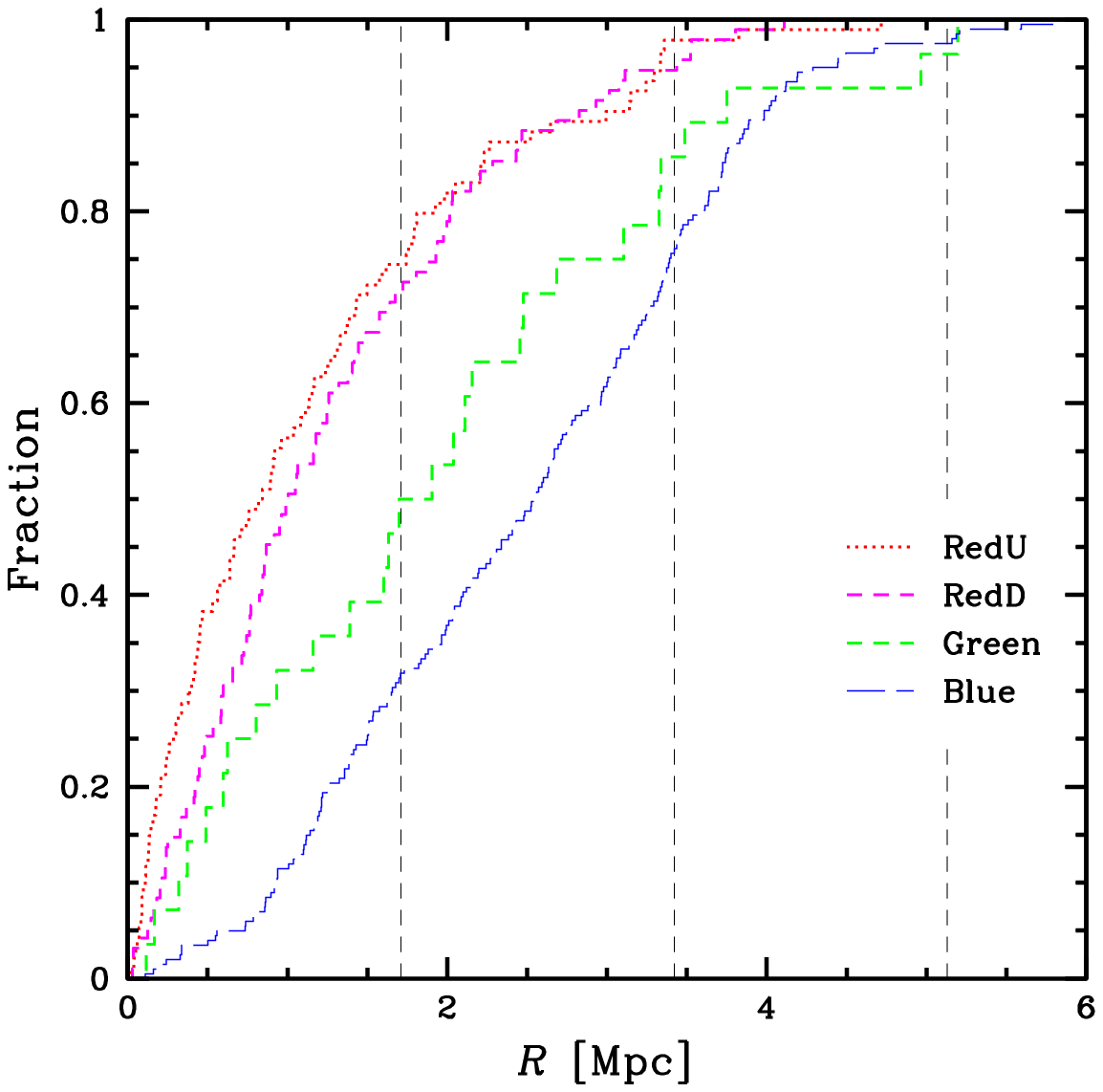}}
\caption
{Cumulative distributions of the clustercentric distance $R$ of galaxies
  per color-type class (see Fig.~\ref{figcm}). Vertical dashed lines indicate the positions of one, two, and three $R_{200}$.  
}
\label{figKWr}
\end{figure}

The application of the KW test to the rest frame velocities $V_{\rm
  rf}$ [i.e., $(V-\left<V\right>)/(1+z)$] does not give a significant
result, that is the velocity distributions of different color type
classes can originate from the same parent distribution. This is shown
in Fig.~\ref{figKWv} where we also notice a remarkable symmetry in the
distribution of galaxy velocities of all populations around the mean
cluster redshift. This symmetry is quantified measuring the skewness indicator
in Sect.~\ref{sub}. We obtain no significant differences when comparing
the distributions of $|V_{\rm rf}|$, which is consistent with the fact
that there is no significant difference among the velocity dispersions
of the different populations (see Tab~\ref{tabvv}).

Figure~\ref{figvdprof} shows the integral profiles of the mean
velocity and velocity dispersion of red and blue galaxies, separately.
The global mean velocities of the two populations are consistent
between them and with the BCG velocity within the measurement
uncertainties.  The integral profile of the velocity dispersion of the
blue galaxies shows a stronger decrease than that of the red
galaxies. Figure~\ref{figvdprof} (bottom panel) also shows the
differential profiles.

\begin{table}
        \caption[]{Kinematical  properties of the whole cluster and galaxy populations.}
         \label{tabvv}
                $$
         \begin{array}{l r l l}
            \hline
            \noalign{\smallskip}
            \hline
            \noalign{\smallskip}
\mathrm{Sample} & \mathrm{N_g} & \phantom{249}<V>\phantom{249} & 
\phantom{24}\sigma_V\phantom{24}\\
& &\phantom{249}\mathrm{km\ s^{-1}}\phantom{249}&\phantom{2}\mathrm{km\ s^{-1}}\phantom{24}\\
            \hline
            \noalign{\smallskip}
\mathrm{TOT}        &430&135011\pm40 &\phantom{1}841_{-36}^{+26}\\
\mathrm{R200 }      &227&135042\pm67 &        1018_{-48}^{+40}\\
\mathrm{Red  }      &189&135020\pm64 &\phantom{1}884_{-52}^{+40}\\
\mathrm{RedU  }     & 94&135070\pm91 &\phantom{1}880_{-61}^{+59}\\
\mathrm{RedD  }     & 95&134969\pm92 &\phantom{1}892_{-64}^{+69}\\
\mathrm{Green }     & 28&134773\pm163&\phantom{1}842_{-71}^{+102}\\
\mathrm{Blue }      &201&135007\pm55 &\phantom{1}783_{-44}^{+40}\\
              \noalign{\smallskip}
            \hline
         \end{array}
$$ \tablefoot{Col.~1: sample ID; Col.~2: the number of assigned
           galaxies, $N_{\rm g}$; Cols.~3 and 4: mean LOS velocity and
           velocity dispersion of galaxies, $\left<V\right>$ and
           $\sigma_V$, with respective errors.  }
         \end{table}


\begin{figure}
\centering
\resizebox{\hsize}{!}{\includegraphics{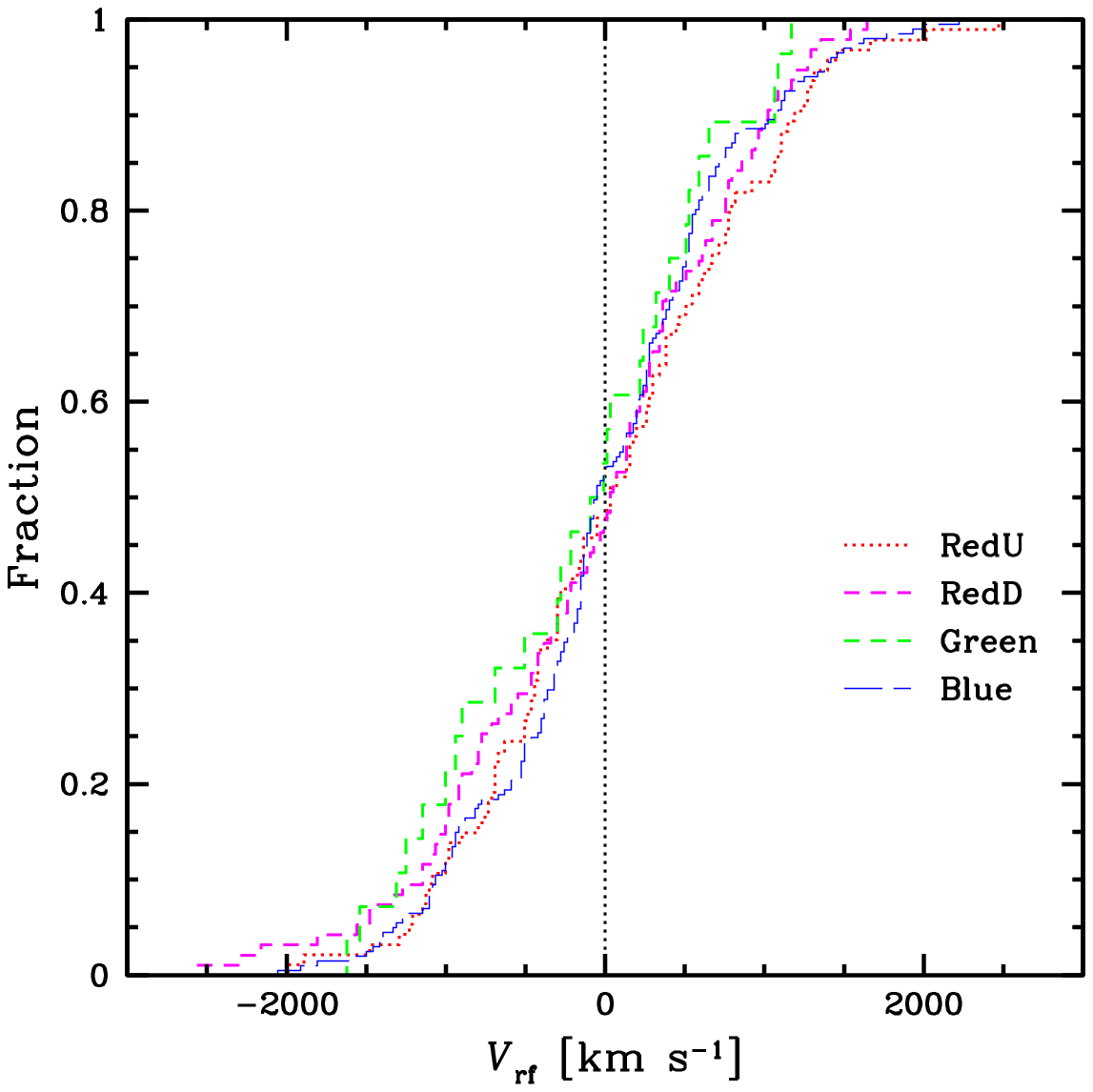}}
\caption
{Cumulative distributions of the rest-frame velocity $V_{\rm rf}$ 
 of galaxies per color-type class. The dotted line corresponds to the mean cluster redshift.}
\label{figKWv}
\end{figure}

\section{Substructure}
\label{sub}

Here we present our analyses and results on the features that indicate
a deviation of the cluster from a dynamical equilibrium state and can
be used as tracers for the cluster formation (see, e.g.,
\citealt{girardi2002} for a review).  Since the sensitivity of any
substructure diagnostics is known to depend on the relative position
and velocity of the substructure, we performed a series of tests in
velocity space (1D tests), in the space of positions projected onto
the sky (2D tests), and in combined space (3D
tests). Table~\ref{tabsub} summarizes our main results on the
substructure.

\begin{table*}
        \caption[]{Results of the substructure analysis.}
         \label{tabsub}
                $$
         \begin{array}{l r |  c c c c c | c | c c }
            \hline
            \noalign{\smallskip}
            \hline
            \noalign{\smallskip}
&
&\multicolumn{5}{c|}{\mathrm{1D}} 
&\multicolumn{1}{c|}{\mathrm{2D}}
&\multicolumn{2}{c}{\mathrm{2D}+\mathrm{1D}}\\
\mathrm{Sample} & N_{\rm g}
&\mathrm{S}&\mathrm{K}& \mathrm{TI}&\mathrm{AI}&\mathrm{Vpec}
&\mathrm{DED.} 
&\mathrm{Vgrad.}&\mathrm{DSV}\\
&
& \%&\%&\%&\%&\%
& \mathrm{N_p}
&\%& \% \\
            \hline
            \noalign{\smallskip}
\mathrm{TOT}     &430&ns&ns &90{\rm -}95 &ns&ns   &1+2     &\sim 94&99.8  \\
\mathrm{R200 }   &227&ns&ns        &ns    &ns   &ns&1+2     &ns&ns  \\
\mathrm{Red  }   &189&ns& ns       &ns    &ns   &-&1+2     &ns&ns  \\
\mathrm{Green }  & 28&ns&90{\rm -}95&ns    &ns&-   &1       &ns&ns  \\
\mathrm{Blue}    &201&ns&ns   &95{\rm -}99 &95{\rm -}99&-&2       &ns&99.3  \\
              \noalign{\smallskip}
            \hline
         \end{array}
$$ \tablefoot{Col.~1: sample ID; Col.~2: number of galaxies, $N_{\rm
             g}$; Cols.~3, 4, 5, 6: significance of the deviations
           from Gaussian according to the skewness (S), kurtosis (K), tail,
           and asymmetry indices (TI and AI); Col.~7: significance of the
           peculiarity of the BCG velocity; Col.~8: number of peaks
           detected through the 2D-DEDICA method, where n1+n2
           indicates the presence of n1 major peaks and n2 minor with
           very low-density peaks; Col.~9: significance of the
           existence of a velocity gradient; Col.~10: significance of
           the existence of substructure according to the DSV test.
           Only significance values larger than $90\%$ are reported,
           while non-significant values are indicated with 'ns'.}
         \end{table*}


\subsection{Analysis of the velocity distribution}
\label{1D}

The velocity distribution was analyzed for possible deviations from
the Gaussian distribution, which could provide important information
about the complex internal dynamics. In fact, a Gaussian
  distribution is assumed for the velocities of galaxies in clusters
  because clusters are expected to relax to an equilibrium
  configuration. The inspection of the phase space distribution of
  galaxies in Fig.~\ref{figvd} suggests a certain
  degree of persistent infall
  onto MACS0329 and it is known that the extent of accretion is likely related
  with the Gaussianity or non-Gaussianity of the velocity distribution
  (e.g., \citealt{sampaio2021}). However, it is expected that the infall
  is more important for external cluster regions and blue galaxies. From
  this perspective, we analyzed several subsamples of the cluster
  data. The results of our analyses on the cluster substructure are
  discussed in Sect.~\ref{discu}.

  We calculated the shape
estimators proposed by Bird \& Beers (\citeyear{bird1993}; see their
Table~2), that is the skewness (SKEW), the kurtosis (KURT), the
asymmetry index (AI) and the tail index (TI).  We found marginal
evidence of non-Gaussianity for the whole sample (lightly tailed
according to TI, in the range of 90--95\% c.l.) and no evidence for
the sample of galaxies within $R_{200}$. Looking at the different
galaxy populations, we found no evidence of substructure for the Red
sample, a marginal evidence for the Green sample (platokurtic in the
range of 90--95\% c.l.), and a significant evidence for the Blue sample
(lightly tailed according to TI and with positive AI, both in the
range of 95--99\% c.l.).

Based on the Indicator test of Gebhardt \& Beers
(\citealt{gebhardt1991}), we also checked the peculiarity of the BCG
velocity ($V_{\rm BCG}=135\,027$ \kss) in the two samples containing
the BCG (namely the TOT and R200 samples). There is no indication of
an anomaly (see also Fig.~\ref{figvdprof}, middle panel).  To detect
and analyze possible deviations from a single-peak distribution, we
also applied the 1D-DEDICA method previously used  in Sect.~\ref{memb} to
determine MACS0329 membership. For each sample, there is no evidence of
a multimodal distribution.

\subsection{Analysis of the 2D galaxy distribution}
\label{2D}

We calculated the ellipticity ($\epsilon=1-b/a$, where $a$ and $b$ are
the major and minor axes) and the position angle of the major axis
(PA) of the galaxy distribution using the method of moments of inertia
(\citealt{carter1980}; see also \citealt{plionis2002} with weight
$w=1$). Here PA are measured counterclockwise from north.
Figure~\ref{figell} shows the integral estimates of $\epsilon$ and PA
at increasing radii out to 4 Mpc, where most member galaxies are
contained.  The results within 0.4 \h are too noisy to be meaningful.
Table~\ref{tabell} lists the values of $\epsilon$ and PA for the
galaxies within 4 Mpc and R200, and for the red and blue galaxies
separately.  The red and blue populations show a clear dicothomy.  As
for the blue galaxies, their distribution is consistent with a round
distribution out to about to $\sim 3$ Mpc, and only in the outermost region
we detect an ellipticity $\sim 0.25$ with a PA consistent with a
north--south elongation. The distribution of the red galaxies is far
from being round, reaching a value of $\epsilon=0.39_{-0.06}^{+0.05}$
at $R_{200}$ and similar values out to 3 Mpc.
For the red galaxies, the PA is therefore already fully significant in
the internal cluster region with PA$=-42_{-5}^{+4}$ degrees at
$R_{200}$ and values between $-40$\degree and $-30$\degree in the entire
investigated region, that is a SE-NW elongation.  The values of
ellipticity and PA of the TOT sample can be explained by the
combination of the two populations, with the blue galaxies dominating
in the outer cluster regions.

\begin{table}
        \caption[]{Ellipticity and position angle of the galaxy distribution.}
         \label{tabell}
                $$
         \begin{array}{l r c r}
            \hline
            \noalign{\smallskip}
            \hline
            \noalign{\smallskip}
\mathrm{Sample} & N_{\rm g} & \epsilon&\mathrm{PA}\\
                &   &      &\mathrm{deg}\\
            \hline
            \noalign{\smallskip}
\mathrm{R<4 Mpc}       &407&0.17_{-0.06}^{+0.04}&3_{-3\phantom{1}}^{+1\phantom{1}}\\
\mathrm{R<R200 }      &227&0.24_{-0.06}^{+0.04}&-38_{-8\phantom{1}}^{+6\phantom{1}}\\
\mathrm{R<R200,Red}   &138&0.39_{-0.06}^{+0.05}&-42_{-5\phantom{1}}^{+4\phantom{1}}\\  
\mathrm{R<R200,Blue}  & 64&0.08_{-0.08}^{+0.01}&-174_{-6\phantom{1}}^{+1\phantom{1}}\\
              \noalign{\smallskip}
            \hline
         \end{array}
$$
\end{table}

\begin{figure}
\centering
\resizebox{\hsize}{!}{\includegraphics{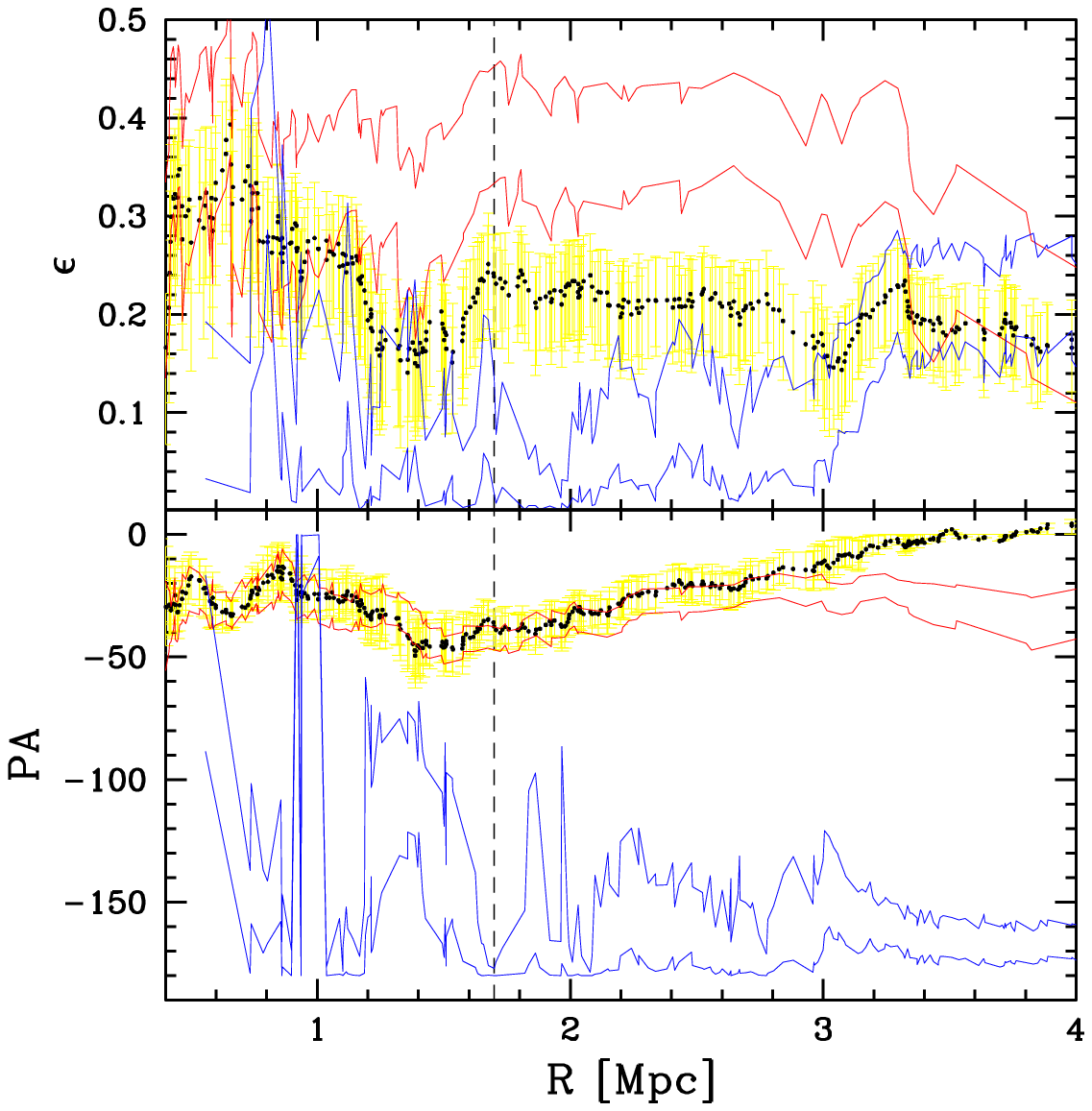}}
\caption
{Integral profiles of ellipticity ({\em upper panel}) and
  position angle ({\em lower panel}) for the entire galaxy population
  (solid black line and yellow errorbars) and for red and blue galaxy
  populations (red and blue errorbands only).  The values of $\epsilon$
  and PA at a given radius $R$ are estimated by considering all
  galaxies within $R$.
  The vertical dashed line shows the value of
  $R_{200}$ (1.71 Mpc).}
\label{figell}
\end{figure}

\begin{figure}
\centering
\resizebox{\hsize}{!}{\includegraphics{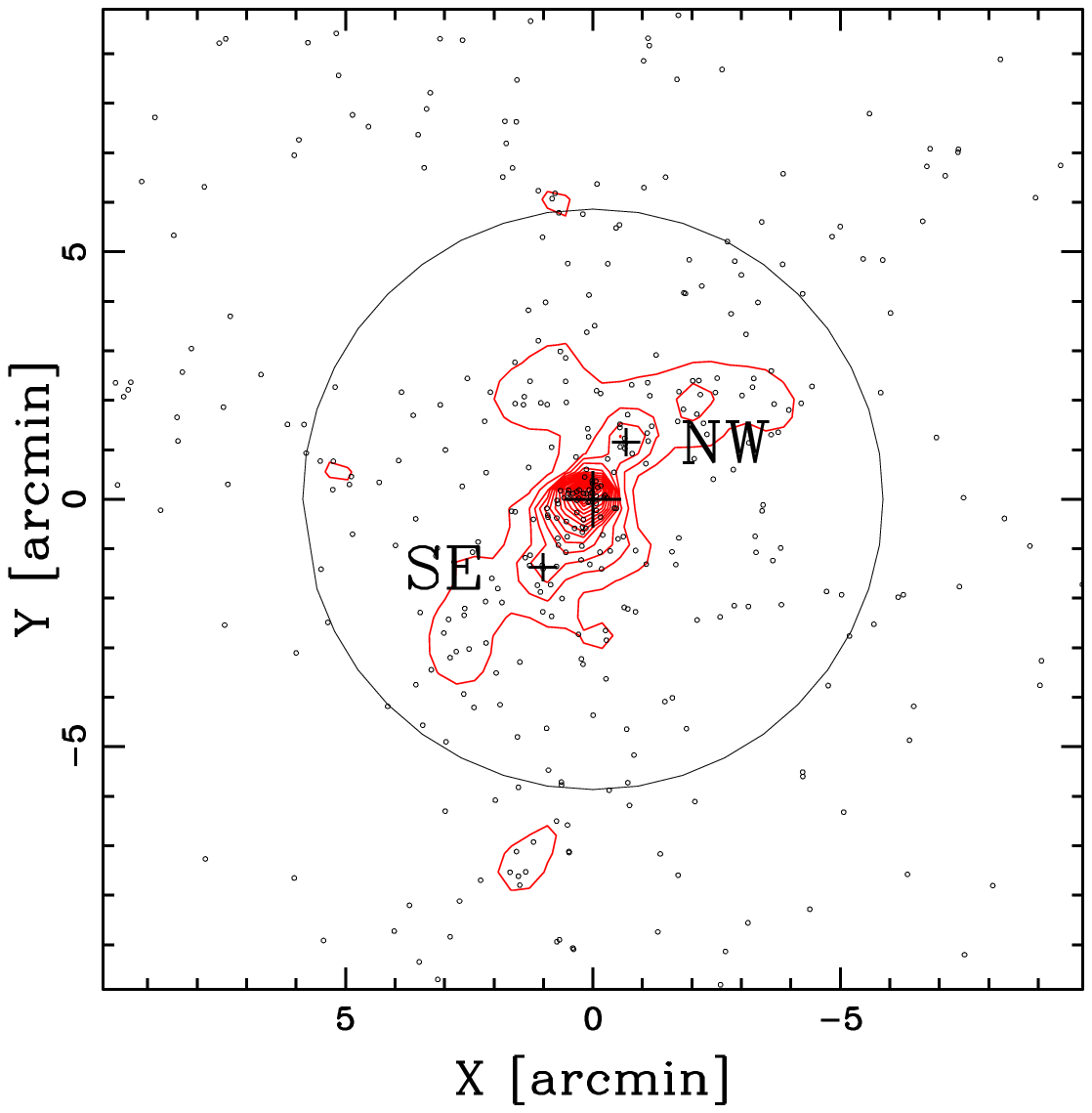}}
\caption{Spatial distribution and isodensity contours of the cluster
  members with $R_{\rm C}\le 24$ within the $20\arcmm \times 20\arcm$
  region centered on the BCG, corresponding to $\sim 7 \times 7$
  Mpc$^2$. Large and small crosses show the position of the BCG and the
  secondary density peaks (see Table~\ref{tabdedica2d}). The $R_{200}$
  region is highlighted by the circle.  }
\label{figk2z}
\end{figure}

\begin{table}
        \caption[]{Results of the 2D-DEDICA analysis.}
         \label{tabdedica2d}
            $$
         \begin{array}{l r c c }
            \hline
            \noalign{\smallskip}
            \hline
            \noalign{\smallskip}
\mathrm{Subclump} & N_{\rm g} & \alpha({\rm J}2000),\,\delta({\rm J}2000)&\rho\\
& & \mathrm{h:m:s,\degree:\arcmm:\arcs}&\\
         \hline
         \noalign{\smallskip}
\mathrm{TOT-main}   &74&03\ 29\ 41.9-02\ 11\ 41&1.00\\
\mathrm{TOT-NW}     &20&03\ 29\ 38.9-02\ 10\ 37&0.21\\
\mathrm{TOT-SE}     &11&03\ 29\ 45.6-02\ 13\ 09&0.17\\ 
         \hline
\mathrm{R200-main}   &75&03\ 29\ 41.9-02\ 11\ 41&1.00\\
\mathrm{R200-NW}     &20&03\ 29\ 39.0-02\ 10\ 36&0.19\\
\mathrm{R200-SE}     &11&03\ 29\ 45.4-02\ 13\ 08&0.16\\ 
         \hline
\mathrm{Red-main}   &58&03\ 29\ 41.8-02\ 11\ 42&1.00\\
\mathrm{Red-NW}     &28&03\ 29\ 39.0-02\ 10\ 40&0.24\\
\mathrm{Red-SE}     &14&03\ 29\ 45.3-02\ 13\ 08&0.20\\ 
         \hline
\mathrm{RedU-main}   &44&03\ 29\ 41.8-02\ 11\ 40&1.00\\
\mathrm{RedU-NW}     &13&03\ 29\ 39.3-02\ 10\ 43&0.17\\
         \hline
\mathrm{RedD-main}   &44&03\ 29\ 42.8-02\ 11\ 41&1.00\\
\mathrm{RedD-SE}     &13&03\ 29\ 45.5-02\ 13\ 13&0.44\\ 
         \hline
\mathrm{Green-main}   &23&03\ 29\ 41.9-02\ 11\ 33&1.00\\
         \hline
\mathrm{Blue-main}   &156&03\ 29\ 46.2-02\ 11\ 31&1.00\\
\mathrm{Blue-South}  & 40&03\ 29\ 41.8-02\ 16\ 04&0.85\\
              \noalign{\smallskip}
            \hline
         \end{array}
         $$ \tablefoot{ Col.~1: subsample/peak ID;
           Col.~2: number of
           assigned member galaxies, $N_{\rm g}$; Col.~3: R.A. and Dec
           of the density peak; Col.~4: relative density with respect
           to the highest peak inside each sample, $\rho$.}

\end{table}


\begin{figure*}
\centering 
\includegraphics[width=4.5cm]{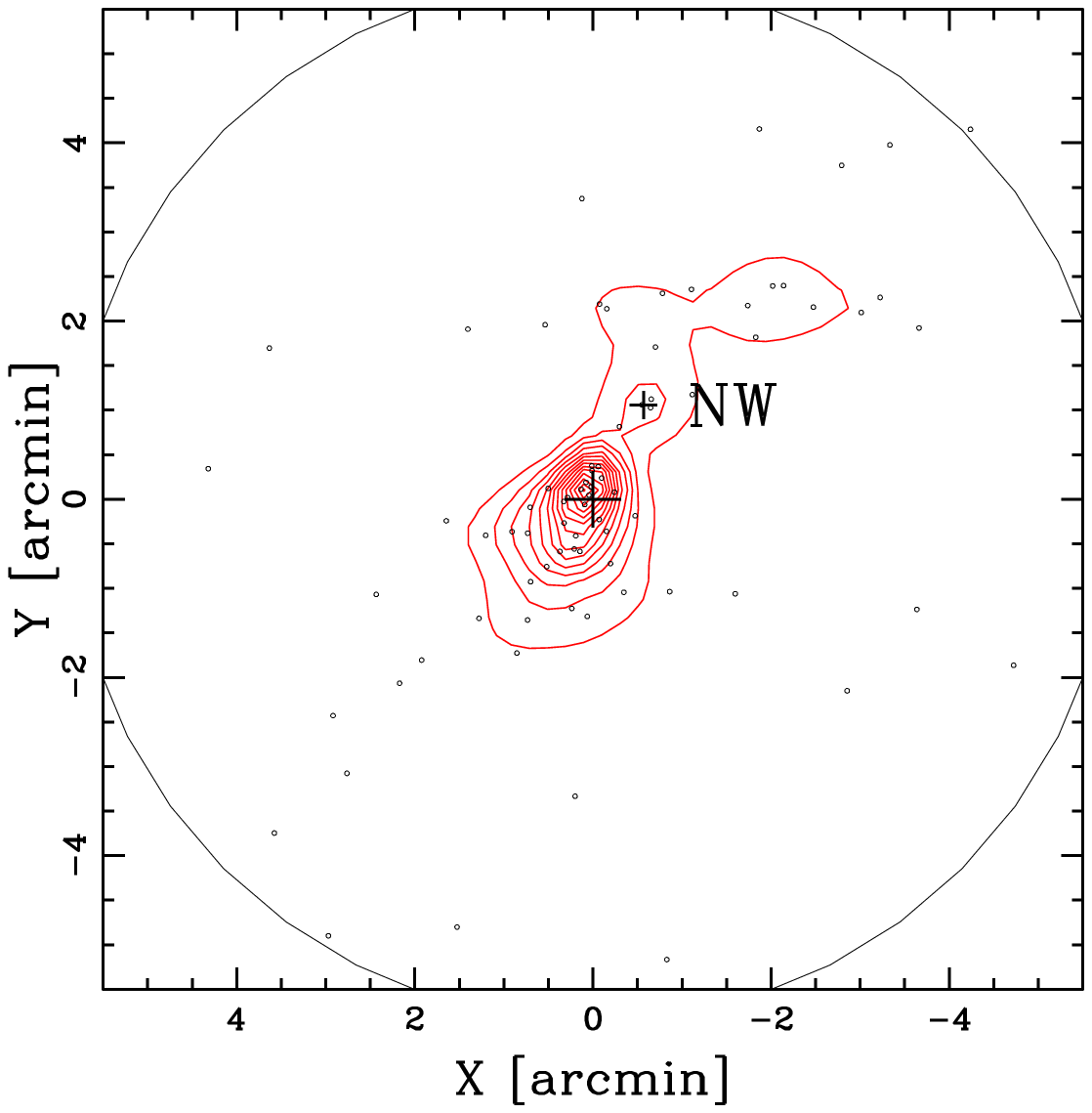}
\includegraphics[width=4.5cm]{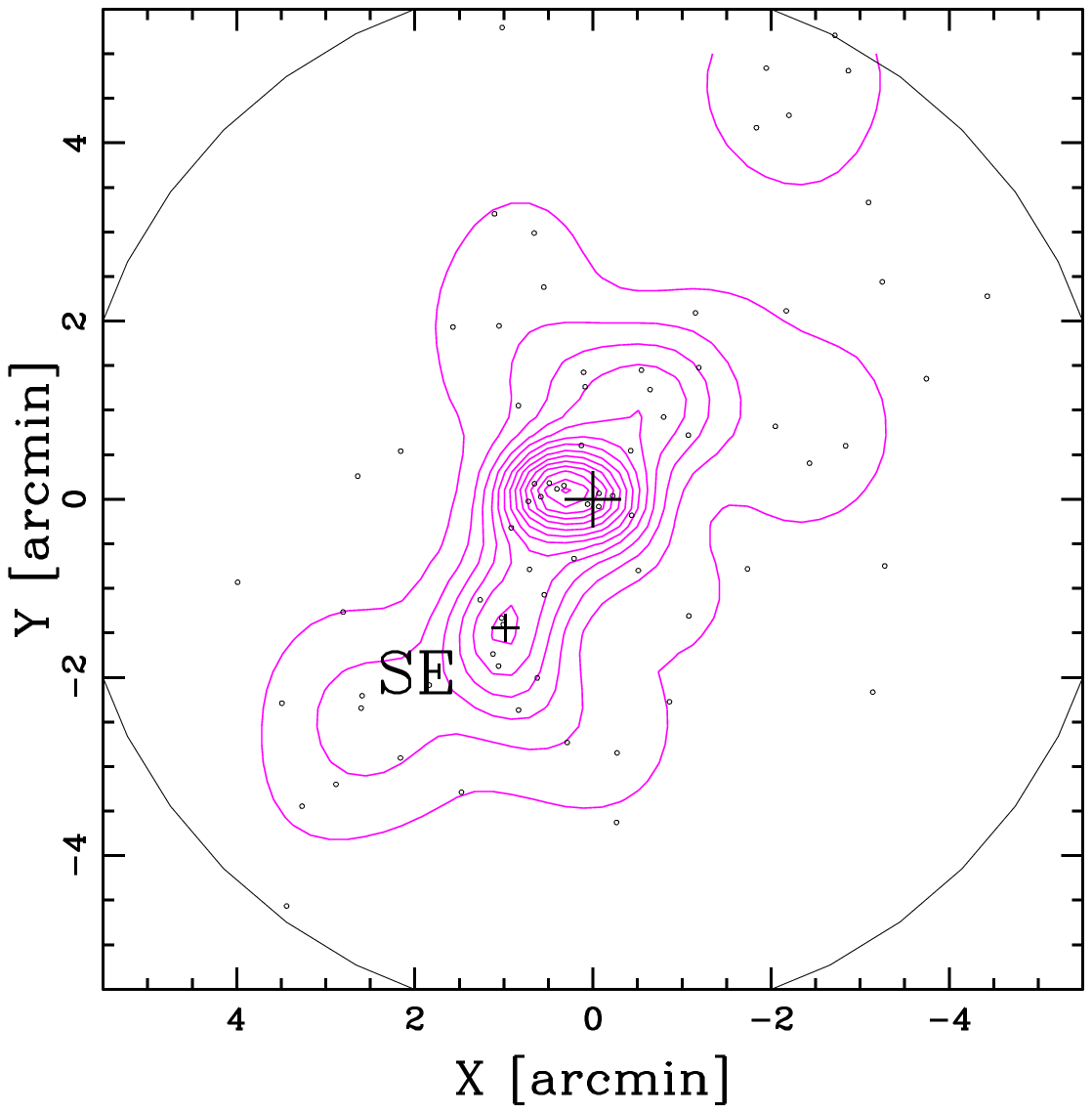}
\includegraphics[width=4.5cm]{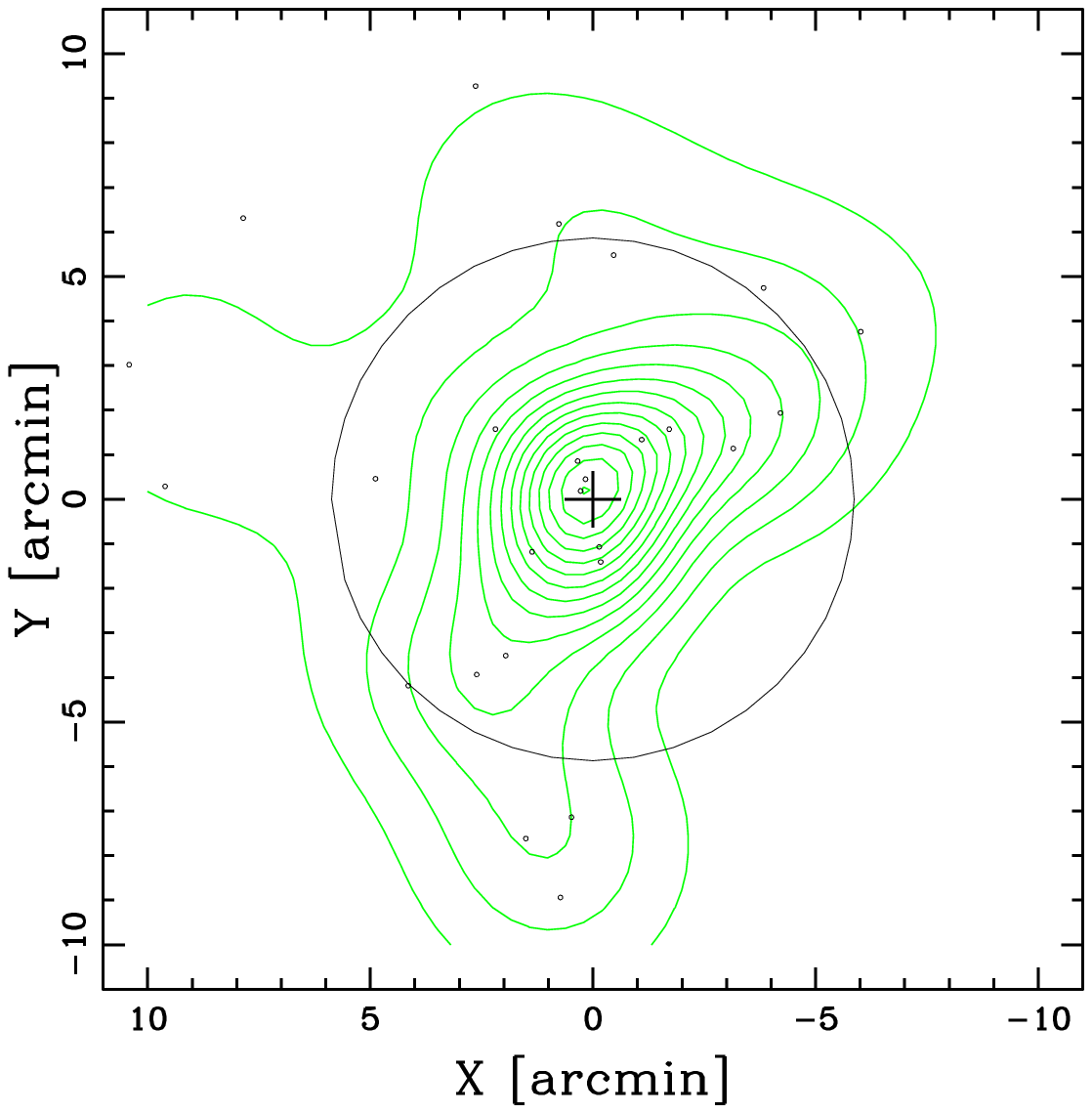}
\includegraphics[width=4.5cm]{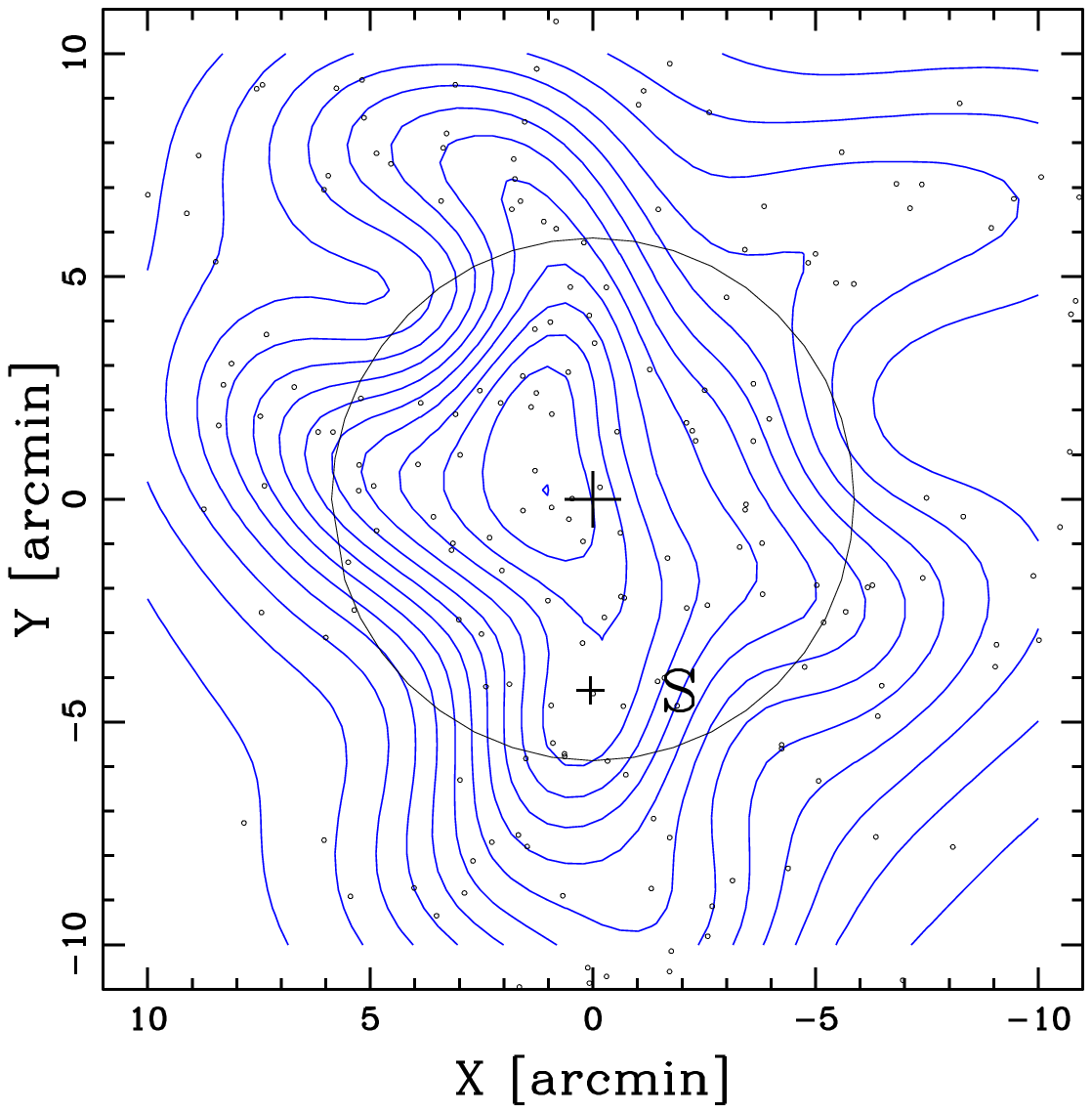}
\caption {Spatial distributions and isodensity contour maps for the
  four color classes: RedU, RedD, Green, and Blue samples (from left to right
  panels). In each panel, we show the region of interest. The
  large and small crosses in each box show the position of the BCG and
  secondary density peaks (see Table~\ref{tabdedica2d}). The $R_{200}$
  region is highlighted by the circle, which is centered on the BCG.}
\label{figk2zcol}
\end{figure*}

We also investigated the spatial distribution of the galaxies using
the 2D-DEDICA method (\citealt{pisani1996}).  To better weight the
substructures with respect to the cluster core, the sample of member
galaxies was restricted to an $R_{\rm C}$-band magnitude of $\le 24$,
that is a sample of 416 galaxies.  This reduces the difference in
spectroscopic completeness between the central region covered by the
MUSE observations and other regions, since $R_{\rm C}=24$ is
approximately the limit reached by the VIMOS redshift
determinations (see also \citealt{biviano2023}).  The only effect in
the color class samples is the rejection of five blue galaxies.

Table~\ref{tabdedica2d} lists the full information for the relevant
peaks, i.e., those with a c.l. $\geq 99\%$, with a relative density
with respect to the main peak $\rho\ \simg 0.20$ and with at least ten
assigned galaxies.  The relevant maps are shown in Figs.~\ref{figk2z}
and \ref{figk2zcol}.  As for the whole cluster, the 2D-DEDICA density
reconstruction shows a main peak centered on the BCG, with two
secondary, smaller peaks in the SE and NW, both within $R_{200}$. The
results for the R200 sample and Red sample are very similar and we do
not show the corresponding figures. We also show the results for RedU
and RedD samples separately, since RedU and RedD galaxies emphasize
the NW peak and the SE peak, respectively (see Fig.~\ref{figk2zcol},
left panels).  The Green and Blue samples appear to be much less
concentrated than the Red sample, although both have a peak near the
BCG.  The spatial distribution of the blue galaxies shows no sign of
SE-NW extension, but a significant southern concentration. This is
consistent with the north--south extension observed in the PA analysis
of the Blue sample when considering external cluster regions.

\subsection{Combining position and velocity information}
\label{3D}

The presence of correlations between positions and velocities of
cluster galaxies is always a strong indication of real substructures.
To investigate the 3D cluster structure we performed two tests.  The
presence of a velocity gradient was searched for by performing a
multiple linear regression fit to the observed velocities with respect
to the galaxy positions in the plane of the sky (e.g.,
\citealt{denhartog1996}; \citealt{girardi1996}).  Significance is
based on 1000 Monte Carlo simulated clusters obtained by shuffling the
velocities of the galaxies with respect to their positions. In the TOT
sample we find a significant velocity gradient consistent with higher
velocities in the south--southeast region than in the north--northwest
region (PA=$170_{-23}^{+26}$, significant at the $94\%$ c.l.).  No
significant patterns are found in other galaxy samples.

To quantify the substructure level, we used our modified version of the $\Delta$ test of Dressler \&
Shectman (\citeyear{dressler1988sub}), which only considers  the
local mean velocity indicator (hereafter DSV test;
\citealt{girardi2010}).
This indicator is
$\delta_{i,V}=[(N_{\rm nn}+1)^{1/2}/\sigma_{V}]\times (\left<V\right>_{\rm loc} -\left< V \right>)$,
where the local mean velocity $\left<V\right>_{\rm loc}$ is calculated
using the $i$-th galaxy and its $N_{\rm{nn}}=10$ neighbors.  For a
cluster, the cumulative deviation is given by the value $\Delta$,
which is the sum of the $\left|\delta_{i,V}\right|$ values of the
individual $N$ galaxies.  As with the velocity
gradient, the significance of $\Delta$ (i.e., the presence of
substructure) is based on the 1000 Monte Carlo simulated clusters.

\begin{figure}
\centering 
\resizebox{\hsize}{!}{\includegraphics{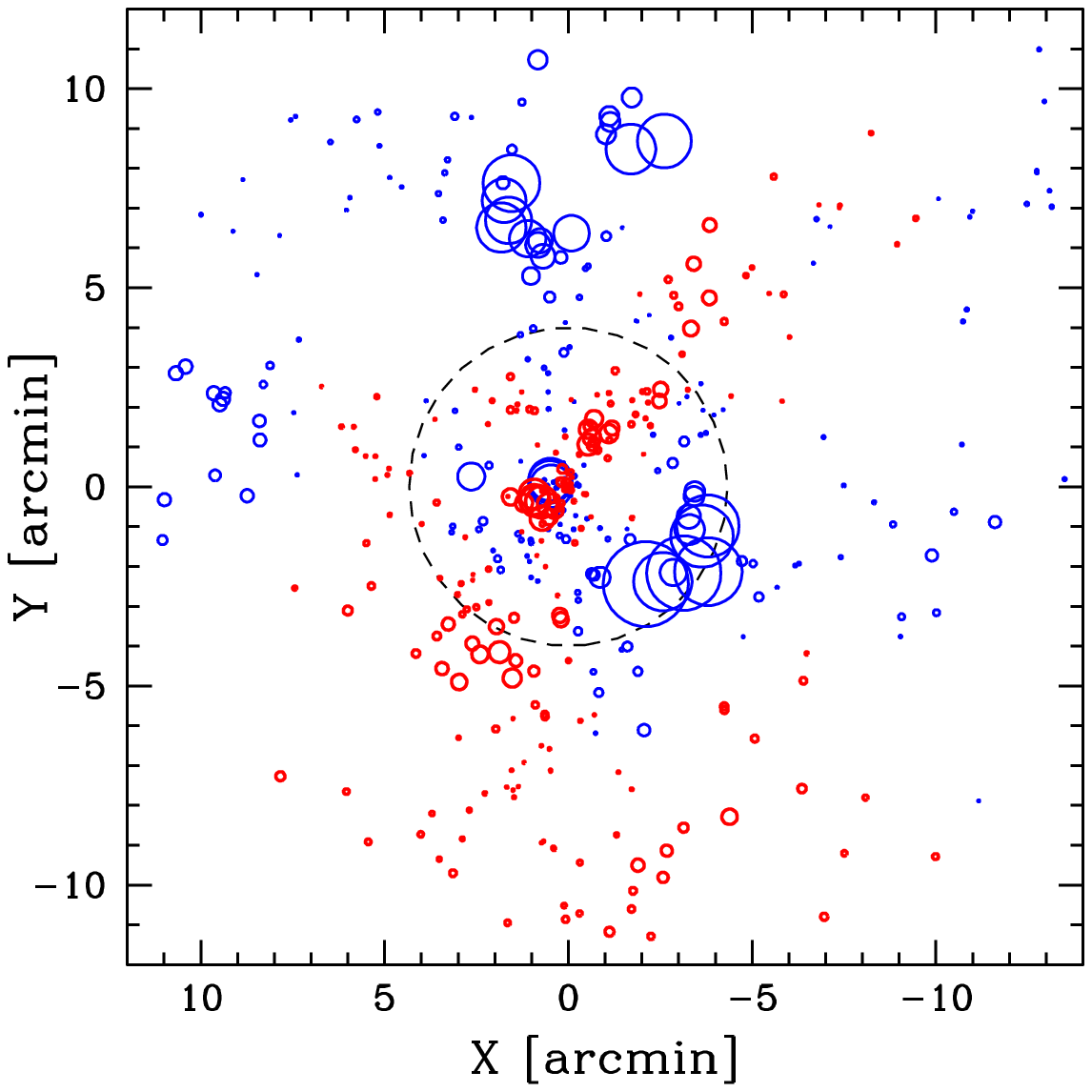}}
\caption{ DSV-test bubble-plot quantifying substructures for the entire cluster population.  The
  larger the circle, the greater the deviation of the local mean
  velocity from the global mean velocity.  Blue and red circles
  indicate where the local value of the mean velocity is smaller or
  larger than the global value.
  The $R_{200}$ region is
  highlighted by the large black dashed circle on the BCG.}
\label{figdssegno10v}
\end{figure}

\begin{figure}
\centering 
\resizebox{\hsize}{!}{\includegraphics{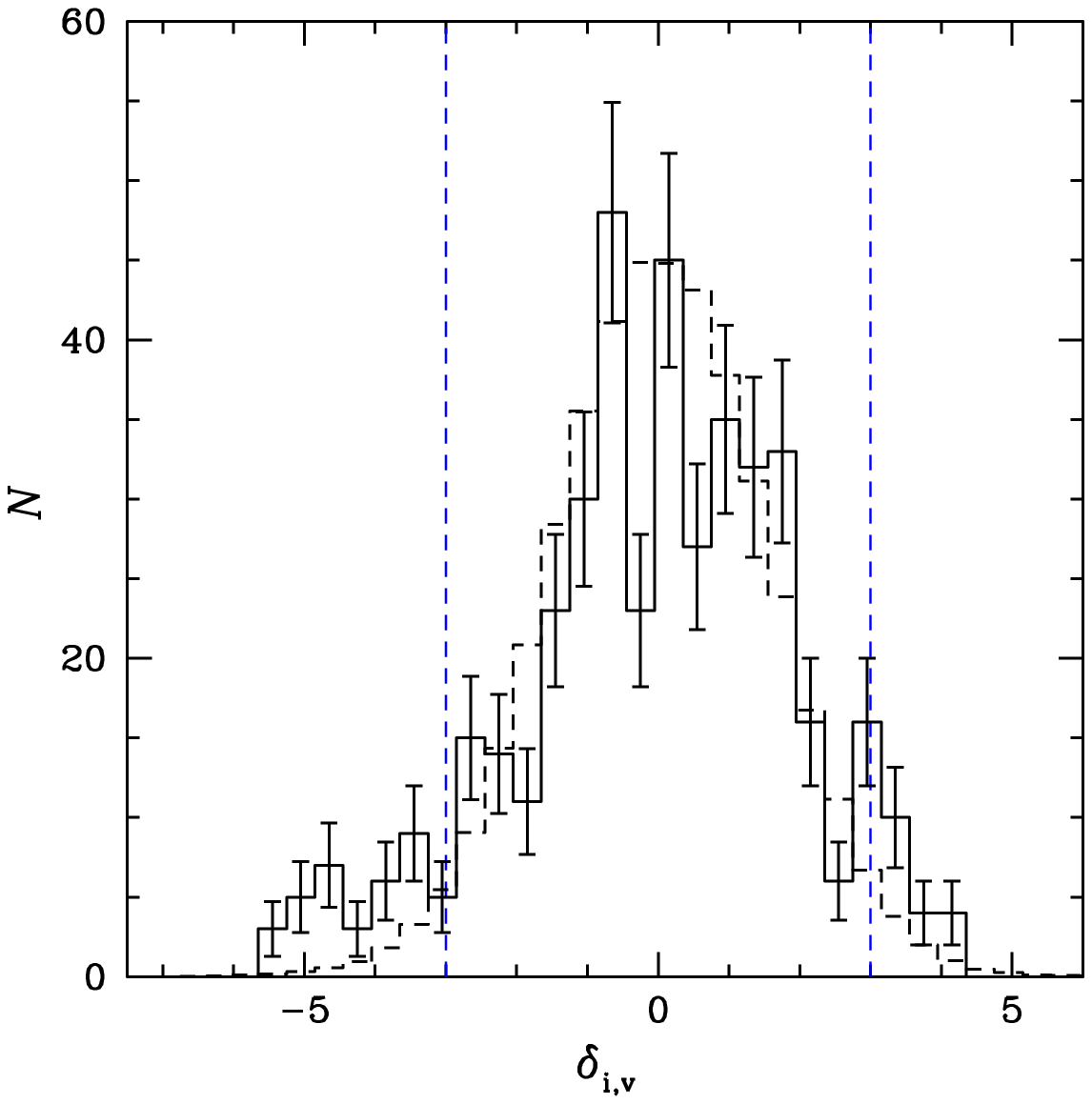}}
\caption
    {Distribution of the $\delta_{i,V}$ values (see text and
      Fig.~\ref{figdssegno10v}). The solid-line histogram 
      refers to the observed galaxies. The dashed-line histogram refers to
      the galaxies of the simulated clusters, normalized to the number of
      observed galaxies. The blue vertical dashed lines mark the
      $|\delta_{i,V}|>3$ regions in which most of the real galaxies
      are expected to be in the 3D substructure.  }
\label{figdeltai}
\end{figure}

The DSV test provides positive evidence of substructure for the entire
population ($99.8\%$ c.l.). The substructure is restricted to two
regions with few low velocity galaxies, as shown in the bubble-plot of
Fig.~\ref{figdssegno10v} (upper panel). The first region is located in the
southwest near $R_{200}$. The second region lies to the
north outside $R_{200}$.  We also used the technique developed by
\citet{biviano2002} to compare the distribution of the $\delta_i$
values of the real galaxies with those of the galaxies of the simulated
clusters.  We compare the distributions of the $\delta_{i,V}$ values
in Fig.~\ref{figdeltai}. The distribution of the real galaxy values shows
a tail at large negative $\delta_{i, V}$ values compared to the
distribution of the simulated galaxy values. This confirms that galaxies
with low velocities are probably the cause of the substructure.
Comparing the two distributions with the 1DKS test, the difference is
quite modest ($\siml 90\%$).  The DSV test finds no substructure in
the R200 sample, the Red sample and the Green sample. The only
exception is the Blue sample.  In this case, the substructure is
significant at the $99.3\%$ c.l., and the bubble-plot again shows a
region populated by low velocity galaxies in the southwest.

All the above analyses suggest that very few galaxies are involved in
the 3D substructures. To verify this, we performed the DSV test
changing the number of neighbors.  The significance of the
substructure decreases when using $N_{\rm{nn}}=20$ instead of
$N_{\rm{nn}}=10$, to $99.4\%$ and $93\%$ for the TOT and Blue samples,
respectively. In contrast, for $N_{\rm{nn}}=5$, the significance of
the substructure increases to $>99.9\%$ and $99.9\%$ c.l. for the TOT
and Blue samples. Figure~\ref{figdssegno5vBlue} shows the bubble plot
for the blue galaxies confirming the presence of a small clump of low
velocity galaxies in the south-southwest.

\begin{figure}
\centering 
\resizebox{\hsize}{!}{\includegraphics{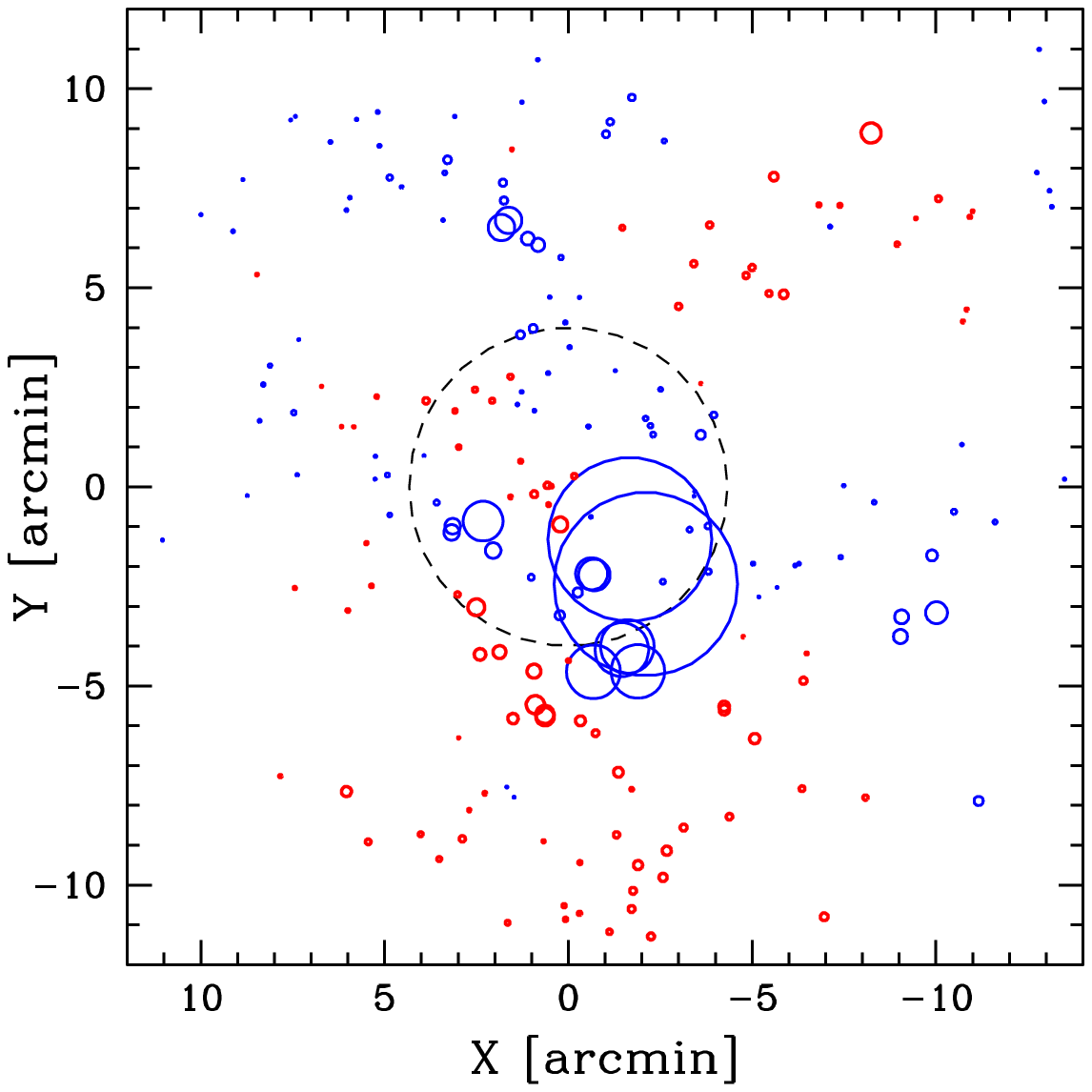}}
\caption
    {As in Fig.~\ref{figdssegno10v}, restricted to the blue galaxies and
      using the DSV-test with
      a smaller number of neighbors to detect small substructures.}
\label{figdssegno5vBlue}
\end{figure}

\section{Exploring galaxy systems in the MACS0329 field}
\label{field}

Here we present our analysis of the other three peaks detected in the
redshift distribution, listed in Table~\ref{tabdedica1d}.  To extract
the member galaxies belonging to each galaxy system (hereafter GrG1,
GrG2, GrG3 in order of increasing redshift), we applied the ``shifting
gapper'' procedure for each peak as already used in Sect.~\ref{memb},
with the standard parameters of Fadda et al. (\citeyear{fadda1996}).
For each galaxy system we show the redshift distribution, the
distribution in phase space and on the sky
(Figs.~\ref{fighistonew}, \ref{figvdpicchi}, and \ref{fig2dpicchi},
respectively).  For each system, details of the membership and
resulting properties are given below and the
global properties are summarized in Table~\ref{tabgroups}.

\begin{table*}
        \caption{Global properties of line-of-sight galaxy systems.}
         \label{tabgroups}
            $$
         \begin{array}{l r c c c c c c r}
            \hline
            \noalign{\smallskip}
            \hline
            \noalign{\smallskip}

\mathrm{Group} &N_{\rm gal} &^{\mathrm{a}}{\rm R.A.(J2000)},\,{\rm Dec.(J2000)}&
            z&\sigma_{V}&N_{R200}&\sigma_{V,200}&R_{200}&M_{200}\\
      &      &\mathrm{h:m:s,\degree:\arcmm:\arcs}&
            &\mathrm{km\ s^{-1}}&&\mathrm{km\ s^{-1}}&\mathrm{Mpc}&10^{13}M_{\odot}\\
         \hline
         \noalign{\smallskip}
\mathrm{GrG1}  &  89&03\ 29\ 33.03-02\ 08\ 51.5&0.3141\pm0.0001&365_{-43}^{+19}&13&338_{-85}^{+75}&0.61\pm0.14&3.6\pm2.5\\
\mathrm{GrG2^{\mathrm{b}}}&  124&03\ 29\ 39.14-02\ 11\ 29.2&0.3848\pm0.0001&352_{-38}^{+16}&27&415_{-63}^{+34}&0.72\pm0.08&6.4\pm2.2\\
\mathrm{GrG3} &  89&03\ 29\ 39.20-02\ 20\ 28.0&0.4730\pm0.0001&432_{-35}^{+31}&25&584_{-75}^{+67}&0.97\pm0.12&17.1\pm6.6\\
              \noalign{\smallskip}
            \hline
         \end{array}
$$
\begin{list}{}{}  
\item[$^{\mathrm{a}}$] As for the center definitions, see text. 
\item[$^{\mathrm{b}}$] Note that the galaxies of GrG2bis are quite sparse
  in the
  field and we do not count GrG2bis among the true, bound groups (see text).
\end{list}
         \end{table*}


\begin{figure}
\centering
\resizebox{\hsize}{!}{\includegraphics{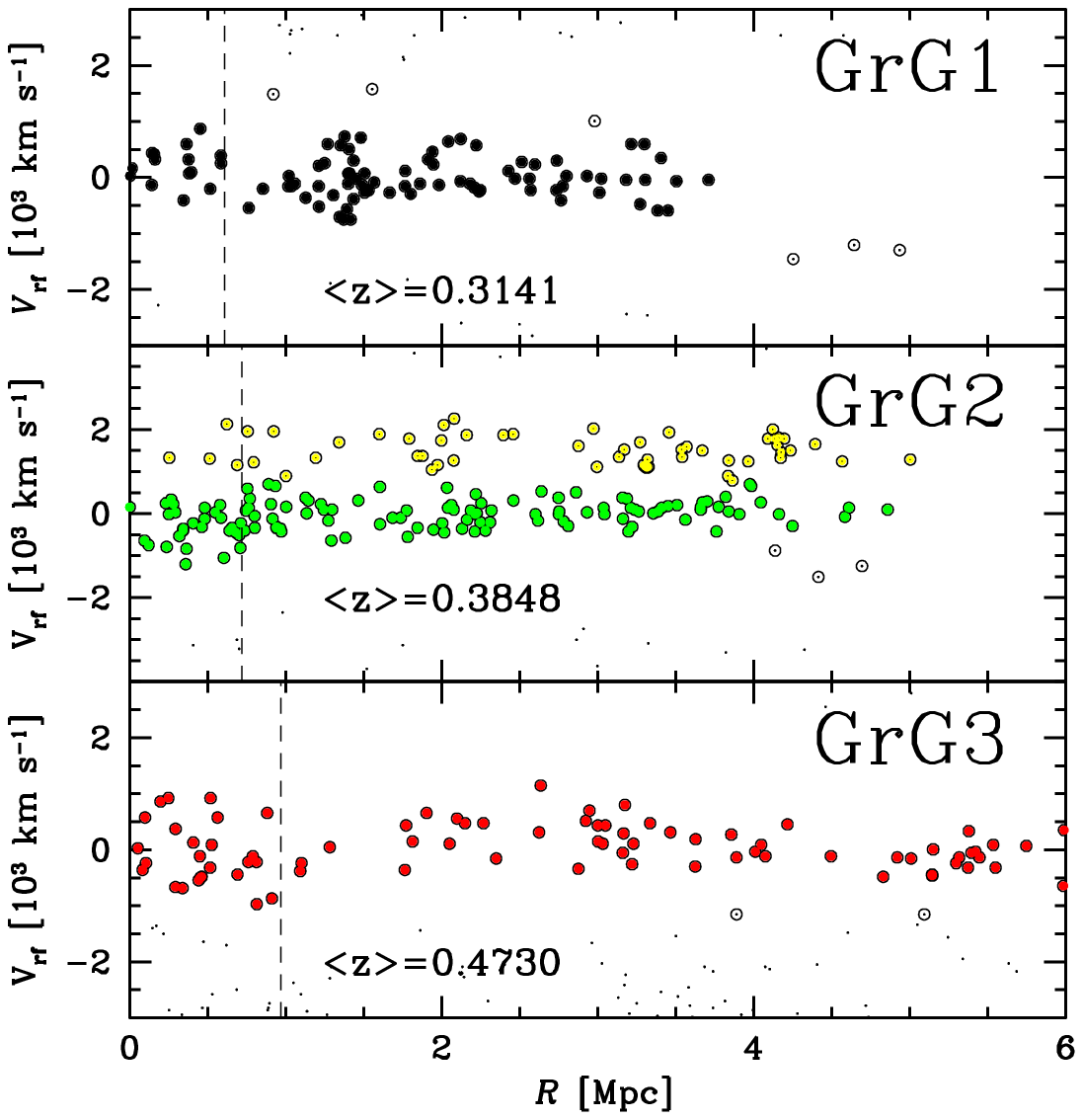}}
\caption
    {Projected phase spaces of the three galaxy systems (GrG1, GrG2,
      and GrG3 in {\em top}, {\em middle}, and {\em bottom panels},
      respectively).  For each system, the rest-frame velocity,
      $V_{rf}$, and the distance $R$ are computed using the group mean
      velocity and group center.
      Circles  indicate galaxies that belong to the velocity
      peak. Full circles indicate the group members (black, green and
      red colors for GrG1, GrG2 and GrG3, respectively).  {\em
        Middle panel} also shows the galaxies of a secondary peak
      (GrG2bis, yellow color). For each group, the vertical line shows
      the $R_{200}$ radius and the mean redshift of the system is
      given (see also Table~\ref{tabgroups}).}
\label{figvdpicchi}
\end{figure}

\begin{figure}
\centering 
\resizebox{\hsize}{!}{\includegraphics{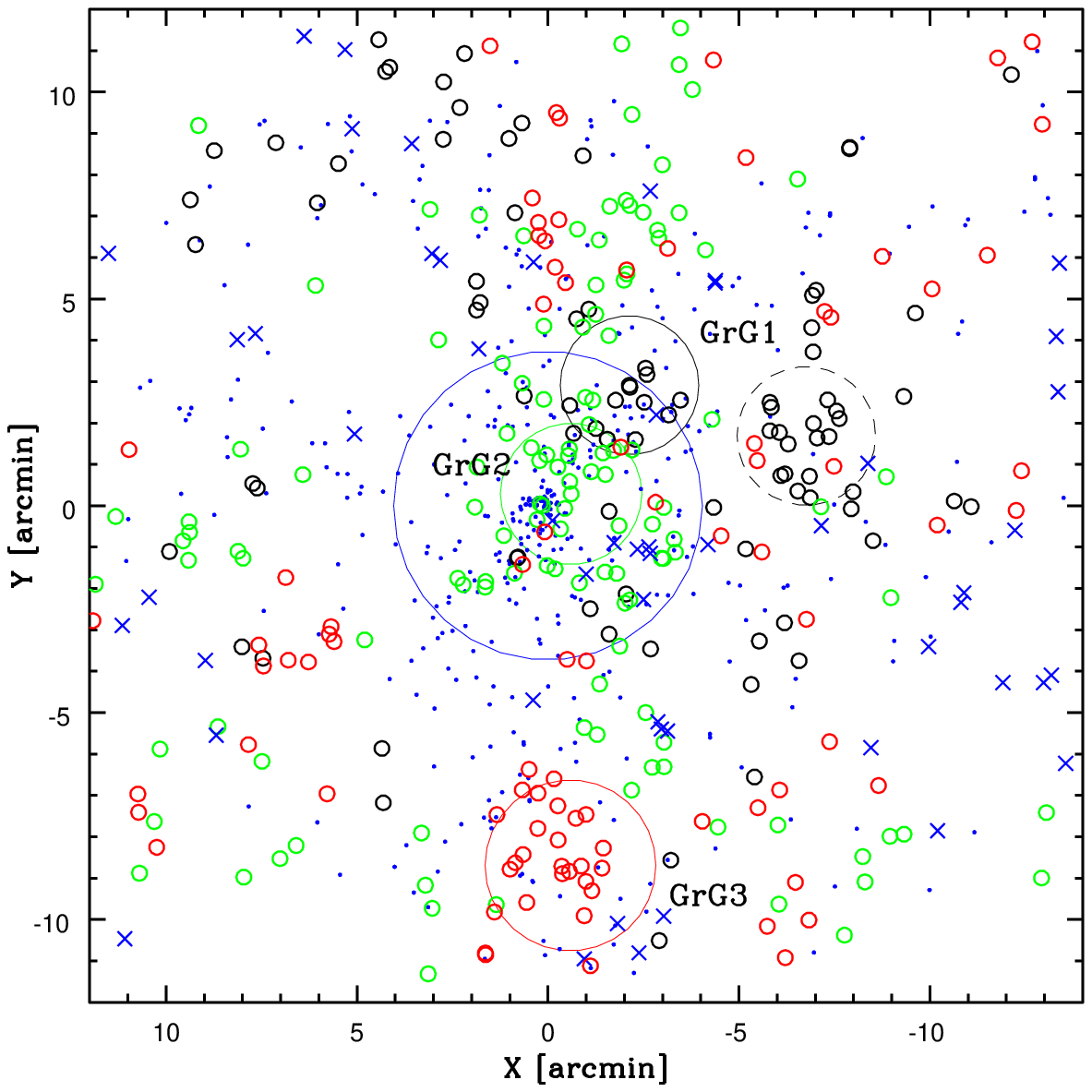}}
\caption
    {Spatial distribution of the 89, 124, and 89 member galaxies of
      GrG1, GrG2, and GrG3 shown with small circles (black, green, and
      red as in Fig.~\ref{figvdpicchi}).  The corresponding $R_{200}$
      radii are represented by large black/green/red circles.  Small
      blue dots and the large blue circle indicate 430 members and the
      $R_{200}$ radius of MACS0329.  The black dashed circle refers to
      the western galaxy concentration of GrG1 (see text).  Galaxies
      belonging to the secondary peak GrG2bis are very sparse in
        the field as shown by blue crosses).  The plot is centered
      on the BCG of MACS0329.}
\label{fig2dpicchi}
\end{figure}

As for the peak no.~1 detected in the redshift distribution, we found
that the 2D spatial distribution of its galaxies shows two
concentrations that are $\sim 5\arcm$ apart (see
Fig.~\ref{fig2dpicchi}).  The eastern concentration is dominated by a
pair of bright galaxies surrounded by diffuse intragroup light (see
Fig.~\ref{figimage}, the GrG1 region).  Unfortunately, we do not have
a redshift measurement for the northeastern galaxy of the pair, but
the photometric redshift is consistent with that of GrG1 ($z_{\rm
  phot}=0.321\pm0.019$ from the Dark Energy Camera Legacy Survey, DESI
DR9). The southwestern galaxy is the second brightest galaxy among the
galaxies in the redshift peak no.~1 ($R_{\rm C}=18.81$, only $\siml
0.5$ fainter than the brightest one).  Therefore, we decided to assume
this galaxy having spectroscopic redshift as the center of the system
[R.A.=$03^{\mathrm{h}}29^{\mathrm{m}}33\dotsec03$, Dec.=$-02\degree
  08\arcmm 51.5\arcs$ (J2000.0)].  The member selection with the
``shifting gapper'' leads to 89 galaxies in the GrG1 structure.  We
calculated the mean redshift and the velocity dispersion.  We used the
same procedure as in Sect.~\ref{prop} to obtain $R_{200}$ and
$M_{200}$. In this way, we identified a probable group with
$R_{200}\ \sim 0.6$ Mpc.  Note that the first brightest galaxy of the
peak no.~1 is located in the western concentration, but at its
edge. This is the reason why we did not analyze this concentration
further. We show this western concentration in Fig.~\ref{fig2dpicchi}
(dashed circle) by using the 2D-DEDICA peak as the center and a radius
with the same size as the GrG1 group.

As for peak no.~2, reapplying the 1D-DEDICA method shows the presence
of two peaks (see Fig.~\ref{fighistonew}, middle panel in the inset
plot).  The low-velocity peak is richer in galaxies and spatially
quite dense around a very luminous dominant galaxy. This galaxy is
projected near the center of MACS0329, $\sim 40$\arcs to the northwest
(see Fig.~\ref{figimage}). Due to its relevance we obtained the
spectrum at TNG (see Fig.~\ref{figspectrum}); other relevant
discussions can be found in Sect.~\ref{dgrou}. It is assumed that this
galaxy is the center of the GrG2 structure
[R.A.=$03^{\mathrm{h}}29^{\mathrm{m}}39\dotsec14$, Dec.=$-02\degree
  11\arcmm 29.1\arcs$ (J2000.0)].  We used the ``shifting gapper''
method to extract 124 members of 127 galaxies in the low velocity
peak. As described above, we calculated the global properties and
extracted a probable group with $R_{200} \sim 0.7$ Mpc.  The galaxies
of the high-velocity peak are quite sparse in the field and we refer
to this structure as GrG2bis  (see blue-cross symbols in
  Fig.~\ref{fig2dpicchi}). They are so sparse in the field that the
  application of the 2D-DEDICA method splits the GrG2bis sample of 55
  galaxies into 34 non-significant subsamples.  The galaxies of GrG2bis could
be part of a large-scale structure or, more interestingly, the
remnants of a previous merger involving GrG2 and now spread across the
field. In any case we no longer discuss GrG2bis.

The galaxies of the peak no.~3 show a dense concentration of galaxies
centered at the southern boundary of the field sampled by our redshift
catalog, $\sim 9\arcm$ south of MACS0329 (see Fig.~\ref{fig2dpicchi}).
There is no clearly dominant galaxy and we assumed that the center is
the density center estimated with the 2D-DEDICA method
[R.A.=$03^{\mathrm{h}}29^{\mathrm{m}}39\dotsec20$, Dec.=$-02\degree
  20\arcmm 28.0\arcs$ (J2000.0)]. The ``shifting gapper'' procedure
leads to 89 galaxies in the structure. As described above, we
calculated the global properties and extracted a probable group with
$R_{200} \sim 1$ Mpc.

We also analyzed the region around MACS0329 using {\em Chandra} X-ray
data and DESI DR9 photometric data.  Figure~\ref{figXD}, that covers
an area of $\sim$30$\times$30 arcmin in the field of MACS0329, shows
the resulting galaxy systems of interest. Considering galaxies with
photometric redshift close to that of MACS0329, i.e. those in the
range 0.41 < $z_{\rm phot}$ < 0.49 and with uncertainty $ez_{\rm phot}$
< 0.05 (blue small symbols in Fig.~18), we applied the Voronoi
Tessellation and Percolation technique (\citealt{ramella2001}; see
also \citealt{barrena2005}; \citealt{boschin2009}).  The galaxies that
we have identified as belonging to structures with a significance at
the 99.9\% c.l. are marked by blue squares.  We are able to recover
both the SE--NW elongation of MACS0329 as well as the galaxy
concentration in the southern regions, that is GrG3.  This result
gives us confidence that the photometric-redshifts can be used to
detect galaxy systems outside the region covered by our spectroscopic
catalog.  We detected another system far to the south [at
  R.A.=$03^{\mathrm{h}}29^{\mathrm{m}}30\dotsec2$, Dec.=$-02\degree
  26\arcmm 24\arcs$ (J2000.0)] and a system to the northeast [at
  R.A.=$03^{\mathrm{h}}30^{\mathrm{m}}45\dotsec1$, Dec.=$-02\degree
  03\arcmm 25\arcs$ (J2000.0)]. The latter is also detected by the
{\em Chandra} X-ray data, as it is an extended emission.  We identify
it with the galaxy system \object{2CXO J033045.8-020323} listed in
NED\footnote{http://ned.ipac.caltech.edu/}. Instead, there is no
corresponding entry in NED for the former galaxy system.

\begin{figure}
\centering 
\resizebox{\hsize}{!}{\includegraphics{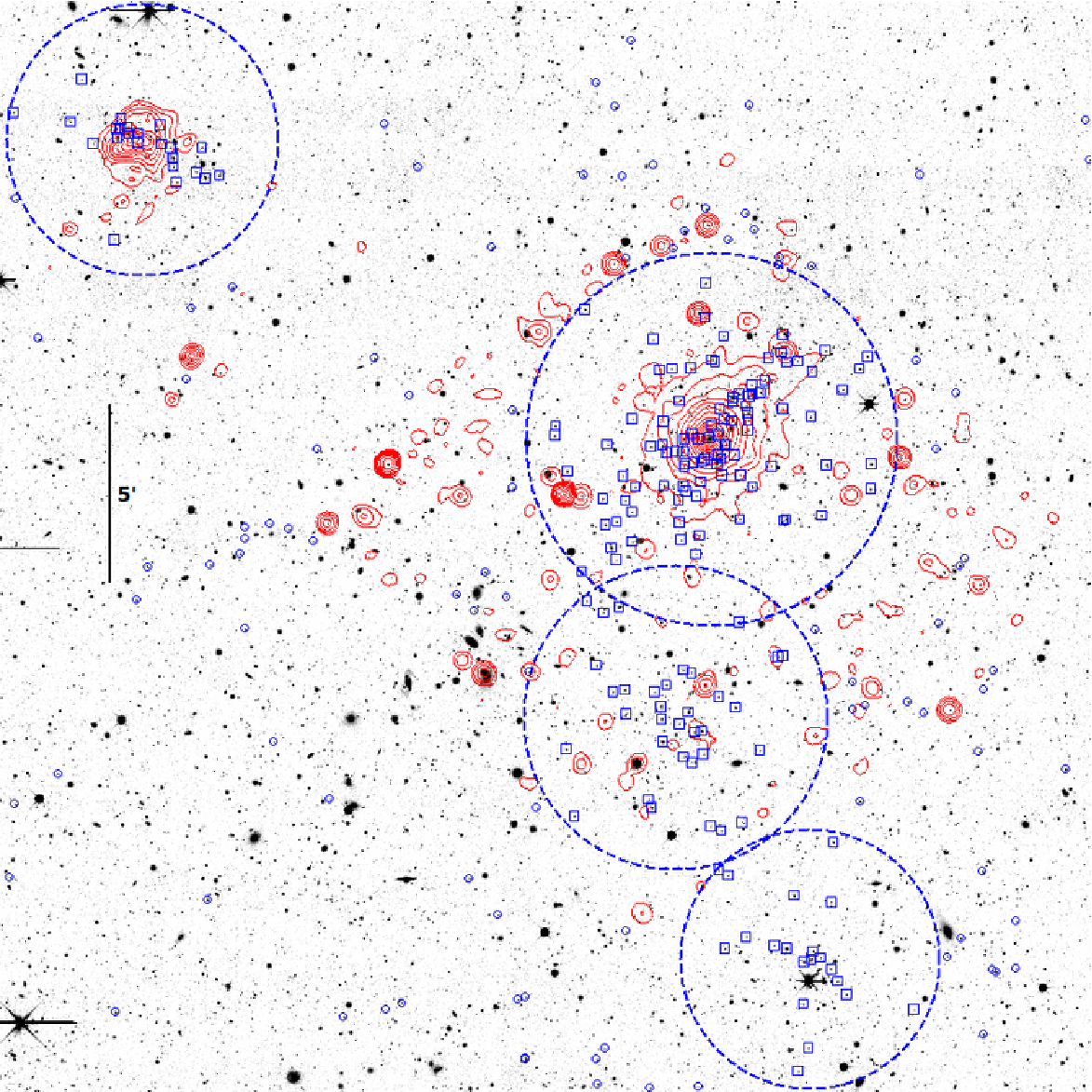}}
\caption{DESI $r$-band image (north top and east left) of the MACS0329 with {\em Chandra} X-ray isocontour levels (in red) superimposed (energy range: 0.5-7 keV).  Small blue symbols indicate galaxies whose
  photometric redshifts are close to that of MACS0329.  In
  particular, blue squares highlight galaxies belonging to significant
  structures according to our Voronoi analysis. Concentrations of
  galaxies are indicated by large blue circles. The largest circle
  indicates MACS0329 with its strong X-ray emission. The galaxy system
  to the south is already detected in the redshift space (GrG3). The
  system further south  lies outside the region covered
  by our redshift catalog. The system in the northeast shows an
  extended X-ray emission and we identify it with 2CXO
  J033045.8-020323.}
\label{figXD}
\end{figure}

\section{Discussion}
\label{discu}

\subsection{Global structure}
\label{dglob}

Our analysis confirms that MACS0329 is a very massive cluster. Our
mass estimate from the scaling relation with the velocity
  dispersion, $M_{200}=(9.2\pm 1.5)$ \mqua agrees with those
  obtained by more sophisticated techniques (MAMPOSSt and Caustic
  methods, see Sect.~\ref{prop}) and fits well within the range 7--13 \mqua of values
published in the literature and derived from X-ray data or
gravitational lensing analyses (\citealt{schmidt2007}; 
  \citealt{donahue2014}; \citealt{umetsu2014}; \citealt{merten2015};
\citealt{umetsu2016}; \citealt{umetsu2018};
\citealt{herbonnet2019}).

The SE-NW elongation of the distribution of galaxies in the plane of sky is
the most important feature of the cluster structure.
This SE-NW direction is essentially traced by the red galaxies and
agrees with the mass distribution from the weak-lensing analysis
(\citealt{donahue2016}; \citealt{umetsu2018}) and the ICM distribution
from X-ray isophotes (\citealt{donahue2016}).  On smaller scales, the
BCG light also follows the same SE-NW direction as shown by the HST UV
and I-bands images (\citealt{donahue2015}; \citealt{durret2019}). The
same direction is also followed by the mass distribution derived from
the strong-lensing analysis (\citealt{okabe2020}).

For the red galaxies within $R_{200}$, we measure a position angle
${\rm PA} = -42_{-5}^{+5}$ degrees and the ellipticity $\epsilon =
0.39_{-0.06}^{+0.05}$.  This ellipticity is higher than the
ellipticity of the X-ray isophotes ($\epsilon \sim 0.15$;
\citealt{maughan2008}; \citealt{mantz2015}), as it is also shown by a direct
visual comparison between the galaxy distribution and X-ray emission
images (cf. Fig.~\ref{figimage} and Fig.~\ref{figk2z}).
The fact that the nearly collisionless dark matter and galaxies
generally have a more elongated distribution than the ICM component is
expected from simulations (\citealt{mcdonald2022}) and confirmed by
observations (\citealt{yuan2023}).

  The velocity dispersion profile of MACS0329 shows the typical
  behavior for clusters, with a modest increase in the central region
  and then a decrease towards external regions.  For instance,
  Fig.~\ref{figvdprof} -- bottom panel -- can be compared to Fig.~2 of
  \citet{girardi1998} obtained by combining 170 clusters.  In the case of
  MACS0329, the extensive redshift catalog allows us to obtain the
  individual profile and to perform a direct comparison with that of
  MACS~J1206.2-0847 which is another CLASH-VLT cluster sampled with
  many redshifts ({\citealt{biviano2023}). Both profiles decline of a
    factor by $\sim 1.5$ between 0.05 and 1 in $R_{200}$ units.

\subsection{Substructure from galaxies vs ICM results}
\label{dsub}

Our study is the first one that uses galaxies to trace the fine
structure of MACS0329.  The statistical results of the substructure
tests applied to the R200 sample show that MACS0329 has no significant
substructure (see Table~\ref{tabsub}) with the exception of two
secondary low-density peaks in the 2D galaxy distribution, which are
caused by the red galaxies. Tests for substructure in velocity space
or in projected phase space only lead to significant results when blue
galaxies or outer cluster regions are considered.  Thus, our analysis
based on member galaxies supports a scenario where MACS0329 is close
to a state of dynamical equilibrium.

Our results agree with \citet{mann2012} who classify MACS0329 among
the relaxed cluster based on the following three criteria: a
pronounced cool core, a perfect alignment of the X-ray peak, and a
single BCG. We note that previous assertions of a non-relaxed status
is due to the analysis of the X-ray morphology and, in particular, to a
few parameters for which MACS0329 is indeed at the borderline between
a status of relaxation or disturbance. According to \citet{mantz2015},
MACS0329 is classified as disturbed because it exceeds the threshold
for the symmetry statistics, but this result is not so conclusive if
we consider the uncertainties related to the measured value
($s=0.85\pm0.08$) compared to the threshold value of $s_{\rm
  th}=0.87$.  According to \citet{maughan2008} and \citet{sayers2013},
MACS0329 is considered a disturbed cluster due to the measure of the
centroid shift ($w=0.014\pm0.003$), which is not significant
when compared to the threshold value $w_{\rm th} = 0.01$.  More
recently, \citet{donahue2016} measure $w=0.011\pm0.001$ for MACS0329
and consider a threshold of $w_{\rm th}=0.02$.  Their Fig.~3 of X-ray
concentration versus the centroid shift shows that MACS0329 resembles
rather a relaxed than a disturbed cluster. Therefore, our results
based on member galaxies well agree with those of \citet{donahue2016}
based on X-ray data.

We think that the only evidence for (a small scale) substructure from
X-ray data comes from the study of \citet{ueda2020} who find a
spiral-like pattern in the X-ray residual image that is consistent
with a gas sloshing core, a phenomenon commonly associated with a
minor merger.  Consistently, MACS0329 shows the presence of a radio
minihalo, which is generally related to a sloshing cold front in cool
cores (\citealt{giacintucci2019}; \citealt{biava2024}).  As evidence
of a past minor merger that caused the gas sloshing,  we propose the two small
substructures that we detect along the SE-NW direction in the 2D
galaxy distribution of all and red galaxies (see, Figs.~\ref{figk2z}
and ~\ref{figk2zcol}).

\subsection{Blue vs red  galaxy populations}
\label{dblu}

We obtain detailed information on the spatial distribution of the red and blue galaxy samples. The
well-known effect of morphological and color segregation in galaxy clusters
(\citealt{dressler1980}; \citealt{whitmore1993}) can be appreciated in Fig.~\ref{figKWr}.  However, higher moments of the phase space distribution of red and blue galaxy populations
can be studied here thanks to the large spatial
coverage and depth of our spectroscopic sample, consisting of 189 and 201
galaxies, respectively, which are used to probe MACS0329 within a radius larger than
$2R_{200}$.

Red galaxies are characterized by a rather elongated distribution,
while the distribution of the blue galaxies is globally round out to
$\sim 3 R_{200}$ (see Fig.~\ref{figell}).  Outside, the distribution of
blue galaxies indicates the north--south direction, and indeed there
is another significant southern concentration at a distance of $\sim
R_{200}$ (see Fig.~\ref{figk2zcol} -- last right panel).  The DSV test
for blue galaxies shows the presence of a galaxy clump in the
south--southwest with a motion that has a component along the
line-of-sight (larger blue circles in Fig.~\ref{figdssegno5vBlue}).
This dichotomy between red and blue galaxies, that is between a more
elongated distribution for red galaxies and a rounder, less
concentrated but more substructured distribution for blue galaxies, is
consistent with the one we have found in two other clusters of the
CLASH-VLT sample studied with extensive redshift catalogs,
MACS~J1206.2-0847 (\citealt{girardi2015}) and  Abell~S1063
(\citealt{mercurio2021}).

  Our interpretation of this dichotomy is the following. Most red
  galaxies are likely tracing the main phase of cluster formation
  along a prominent SE-NW filament of the large-scale structure,
  likely in the plane of the sky since we detect no velocity gradient
  for red galaxies .  It is well known that there is a strong
  correlation between the dynamical activity in galaxy clusters and
  the tendency of the galaxy distribution to be elongated and aligned with
  neighboring clusters (e.g., \citealt{plionis2002}). This is caused by the anisotropic merging
  along the large-scale filaments, as also shown by the analysis of
  simulations (e.g., \citealt{cohen2014}). However, it is also known
  that the final distribution can be aspherical after violent
  relaxation (\citealt{aarseth1978}; \citealt{white1996}).  Therefore,
  the elongation of MACS0329 does not contradict the lack of
  statistical evidence for the presence of an important substructure.
  As for the blue galaxies, our results are consistent with a picture
  in which a cluster irregularly accretes individual spirals or clumps
  of mostly spiral galaxies from the field, that is blue galaxies
  trace the most recent infall into the cluster. In the case of
  MACS0329 we are able to detect two clumps of blue galaxies at a distance
  $\sim R_{200}$. In any case, these clumps of galaxies are very loose
  and should be interpreted as overdensities of the large-scale
  structure filaments that have grown inside the cluster, and not as
  dense and isolated groups.

  In the context of the above scenario, we also discuss the
  kinematical properties of the cluster galaxy populations.  The global
  values of velocity dispersion of the red and blue galaxies agree
  within $\sim 1.3\sigma$ (see Table~\ref{tabvv}) and the agreement is
  also better for galaxies within $R_{200}$. This result supports
  recent  studies indicating  a velocity distribution that does not
  depend dramatically on the color or spectral type of the galaxy
  population (e.g., \citealt{rines2005}; \citealt{mahajan2011};
  \citealt{rines2013}; \citealt{biviano2013}; \citealt{girardi2015}).

    There is evidence that the velocity dispersion profile of
    blue/star forming galaxies decreases with radius more than that of
    red/passive galaxies.  However, this evidence is obtained when
    studying galaxy samples combining many clusters (e.g.,
    \citealt{carlberg1997}; \citealt{biviano1997};
    \citealt{adami1998}; \citealt{dressler1999}).  For
    individual clusters, the spatial morphological segregation and the
    sensitivity of the velocity dispersion measure make more difficult
    to obtain conclusive results.  In the case of MACS0329 we are able
    to see the expected trend (see Fig.~\ref{figvdprof}, bottom
      panel), but the significance is however small.  To highlight
      the difference, we considered the region within 1 Mpc and
      obtained $\sigma_{V,{\rm Blue}}=1368_{-118}^{+119}$ and
      $\sigma_{V,{\rm Red}}=1058_{-65}^{+72}$ for 23 blue and 100 red
      galaxies, resulting in a difference of only 2.2$\sigma$.  Since
      the cluster gravitational potential is the same for all galaxies
      treated as test particles, the sharper decreasing profile is
      interpreted as a sign of more radial orbits in external cluster
      regions (e.g., \citealt{biviano1997}; \citealt{biviano2004};
      \citealt{mamon2019}). For MACS0329, this is consistent with a
      scenario where blue galaxies fell into the cluster more
      recently.

    The history of member galaxies can be traced in a more
    quantitative, statistical way through their position in the
    projected phase space (e.g., \citealt{haines2015};
      \citealt{deshev2020}; \citealt{mercurio2021}; 
      \citealt{qu2023}).  We compare the distribution of galaxies in
    MACS0329 with the halos from the simulations of \citet{oman2013}.
    In Fig.~\ref{figteoOman} we show our data in a way that reproduces
    Fig.~4 of \citet{oman2013}, that is we plot normalized velocities
    $V_{\rm n}=|V_{\rm rf}|/\sigma_{V,{\rm 3D}}$, where we assume
    $\sigma_{V,{\rm 3D}}=\sqrt{3}\times\sigma_{V}$, against normalized
    clustercentric distances $R_{\rm n}=R/R_{\rm vir}$ ($R_{\rm
      vir}=2$ Mpc $\sim 1.2 R_{200}$ for MACS0329). The blue galaxies
    populate the region with large radii and small velocities, that is
    the region of the recently infalled galaxies, while the red
    galaxies populate the region at small radii within a triangular
    region bounded by the line $V_{\rm n}=-(4/3)R_{\rm n}+2$ .  This
    line separates the regions where most galaxies have an infall time
    of less or more than 1 Gyr (\citealt{oman2013}; see also Fig.~1 of
    \citealt{reeves2023}).  In MACS0329, within 1.5 $R_{200}$, only
    9\% of the cluster members can be classified as recently
    accreted. This fraction is smaller than that in the Perseus
    cluster (12\%; \citealt{aguerri2020}) and Abell\,85 (30\%;
    \citealt{agulli2016}).  In agreement with our scenario for
    MACS0329, the fraction of recently accreted galaxies differs for
    blue and red galaxies (18\% and 4\%, respectively).

\begin{figure}
\centering \resizebox{\hsize}{!}{\includegraphics{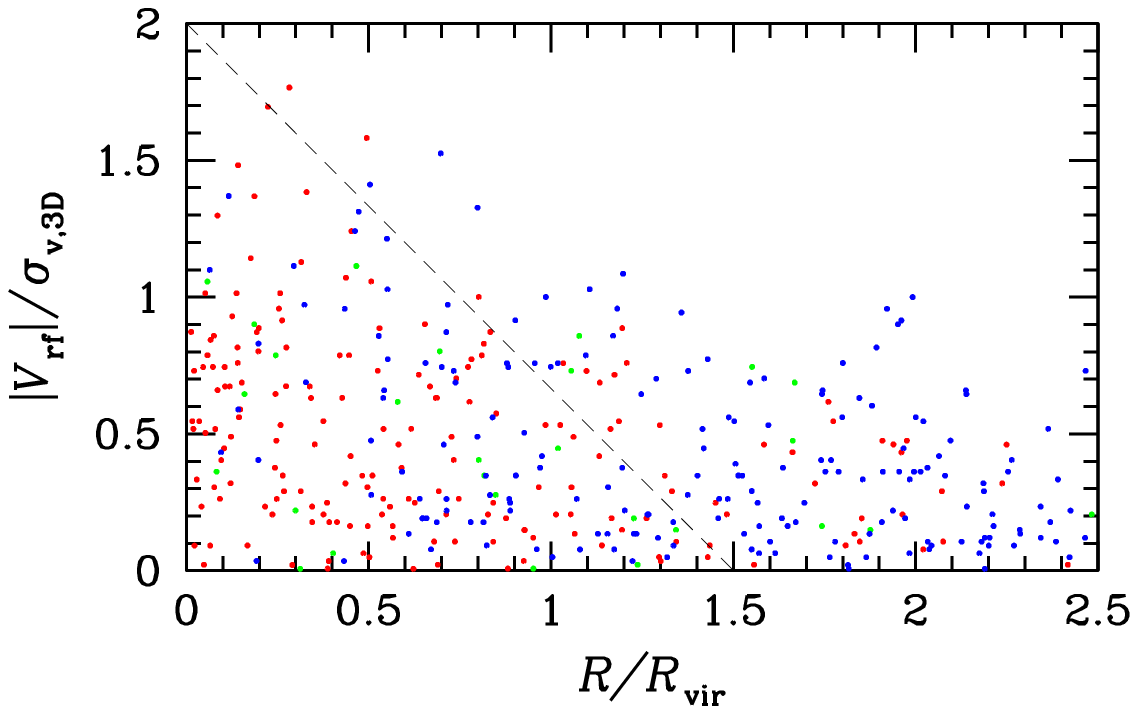}}
\caption{Projected phase space distribution of galaxies per
  different color type (red/green/blue colors for Red/Green/Blue
  samples) showing normalized velocities vs.  normalized projected
  clustercentric distance.  The dashed diagonal line
  separates the regions where most galaxies are expected to have an infall time
  of less or more than 1 Gyr according to the simulations by Oman et al.
  (\citeyear{oman2013}).}
\label{figteoOman}
\end{figure}

\subsection{MACS0329 surroundings and line-of-sight groups}
\label{dgrou}

The analysis of redshift peaks is a very powerful method for detecting
the presence of galaxy systems. For instance, a small peak at $z \sim
0.15$, which is below the relative density threshold adopted in this
study, also proved to be a real group (\citealt{girardi2023}).  Among
the groups that we analyze in this study, GrG3 is the one closest to
MACS0329 in redshift and is the only group that may have an
interaction with MACS0329.  It is also close in projection, less than 3 Mpc in the
cluster rest-frame.  However, by assuming that the redshift difference is
due to kinematics, one obtains that $V_{\rm rf} \sim 4700$ \kss. This
value is larger than the encounter velocity of merging clusters in
cosmological simulations (e.g., \citealt{lokas2023}) and in
observations (e.g., \citealt{sarazin2013}; \citealt{barrena2014};
\citealt{golovich2017}), where 4700 \ks is the maximum observed value
(\citealt{markevitch2006} in the Bullet Cluster). In fact, GrG3 is
gravitationally unbound to MACS0329 according to the simple Newton
criterion when considering only MACS0329 as the total mass of the
system.  However, we could think that both MACS0329 and GrG3 are
embedded in a larger structure such as a supercluster.  In fact,
Figs.~\ref{figvd} and Fig.~\ref{figXD} show that MACS0329 is a cluster
surrounded by many galaxies in the close velocity field and by three
galaxy systems with similar redshift. Unfortunately, two of these
systems are out of our spectroscopic survey and we cannot perform a
more precise analysis.

Finally, we emphasize that an extensive redshift survey is needed to
avoid possible misunderstandings. The brightest galaxy in the GrG2 group
is projected near the center of MACS0329, at $\sim 40$\arcs (see
Fig.~\ref{figimage}) and was therefore wrongly interpreted as a
companion galaxy of the MACS0329 BCG in the context of a post-merger
scenario (\citealt{demaio2015}). Instead, the distance from GrG2 to
MACS0329 in redshift is $\Delta z = 0.0655$, which, if considered of
kinematical nature, corresponds to an enormous difference in velocity
$\Delta V_{\rm rf}\sim 13500$ \kss, which rules out any interaction
between MACS0329 and GrG2 and their respective brightest galaxies.
\citet{ueda2020} had committed the same misunderstanding when they
identified the subcluster responsible for the sloshing core with a
secondary peak in the central mass distribution detected by
\citet{zitrin2015} through strong gravitational analysis (see
Fig.~8 of \citealt{ueda2020}). This secondary peak is located
northwest of the primary X-ray peak and corresponds to the position of
the brightest galaxy in the GrG2 group.  In practice, the secondary
peak of the mass distribution detected by \citet{zitrin2015} is real,
but it is associated with a foreground group rather than a feature of
the internal structure of MACS0329.

An extensive redshift information is also necessary for a correct interpretation of 
results beyond studies of cluster galaxy populations. As example, we note that the visual
inspection of the MACS0329 mass distribution map obtained by
\citet{umetsu2014} through the weak lensing analysis shows the
presence of a second peak $\sim 3.5$\arcm in the
northwest (see their Fig.~1).  The authors do not discuss this feature, however interestingly, this is the
position of a real foreground group (GrG1, see Fig.~\ref{figimage}).
The effect of a small foreground group projected near the cluster
center when deprojecting the mass distribution obtained by
gravitational lensing effects is also discussed for the galaxy cluster
MACS~J1206.2-0847 (\citealt{biviano2023}).  Also in this case, the
detection of the group was feasible thanks to the extensive
CLASH-VLT redshift survey and led to resolve an apparent mismatch between lensing and dynamical mass profiles in the cluster core.

\section{Summary and conclusions}
\label{summa}

We present the first analysis of the massive cluster MACS0329 based on
the kinematics of the member galaxies, as part of the CLASH-VLT ESO
Large Program.  Our analysis is based on an extensive redshift
dataset of over 1700 galaxies, which probe MACS0329 within a radius
of $\sim 3 R_{200}$. The dataset is complemented by multiband photometry
based on high-quality Subaru Suprime-Cam imaging.  We combine the
velocities and positions of the galaxies to select 430 cluster members.  Our
specific results can be summarized as follows.

   \begin{enumerate}
\item
  The new, precise estimate of the cluster redshift is
  $z=0.4503\pm0.0003$.  We report the first estimate of the velocity
  dispersion, $\sigma_{V}=841_{-36}^{+26}$ \kss. Following the
  recipe of \citet{munari2013} and a recursive approach, we compute
  the dynamical mass $M_{200}=(9.2\pm 1.5)$ \mquaa, within a radius
  $R_{200}=(1.71\pm0.07)$ Mpc.  Within this radius, there are 227
  members for which we estimate $\sigma_{V,200}=1018_{-48}^{+40}$
  \kss. 

   \item Using the $B-R_{\rm C}$ vs $R_{\rm C}$ diagram, we classify
     418 member galaxies as blue, intermediate or red galaxies, out to
     $\sim 6$ Mpc. We find a strong spatial segregation among the
     three populations, but no kinematical differences. The only
     exception is the stronger decrease of the velocity dispersion
     profile of the blue galaxies compared to that of the red
     galaxies, corroborating earlier findings reported in the
     literature for combined samples of galaxy clusters and likely due
     to different orbits.
     
     \item Most of the evidence for substructure comes from external cluster
       regions or the blue population. In particular, we detect two
  loose clumps of blue galaxies in the south and southwest at a
  distance $\sim R_{200}$.

       \item The red galaxies show an elongated distribution with an
         ellipticity $\epsilon \sim 0.4$ tracing the SE-NW
         direction. This is consistent with the elongation of the ICM,
         which, however, has a lower ellipticity, as expected given
         that the ICM is a collisional component unlike galaxies. For
         the red galaxies, the only indication of substructure is the
         presence of two minor, secondary peaks in the galaxy
         distribution with a relative density of $\sim 0.2$ with
         respect to the main peak centered on the BCG. This elongated
         SE-NW main structure shows no signs of a velocity gradient,
         which suggests that it lies in the plane of sky.
   \end{enumerate}

       By studying the spatial and velocity distributions of galaxies
       of different colors, we can sketch a probable scenario for the
       assembly history of this massive system. The oldest
       and most important component of MACS0329, consisting of red
       galaxies, traces the main phase of cluster formation along the
       SE-NW direction, while the blue galaxy population shows a more
       rounded distribution and signs of substructure, consistent with
       a more recent and multi-directional infall of groups from the
       field. The comparison with the results of simulations confirms
       different arrival times for red and blue galaxies.  To
       summarize, MACS0329 is not far from dynamical equilibrium. In
       any case, the MACS0329 surroundings are densely populated by  galaxies and three
       close galaxy systems have been detected.

  Our study shows that extensive redshift surveys allow us to
  distinguish galaxy systems along the line of sight and avoid
  misleading results.  In the case of MACS0329, there are two
  foreground groups that mimic a substructure in the projected mass
  map obtained through gravitational lensing analyses.

\section{Data availability}

  Table 1 is only available in electronic form at the CDS via
  anonymous ftp to cdsarc.u-strasbg.fr (130.79.128.5) or via
  http://cdsweb.u-strasbg.fr/cgi-bin/qcat?J/A+A/.

\begin{acknowledgements}

We thank the referee for his/her useful and constructive comments.
  We acknowledge ﬁnancial contributions by the grant MIUR PRIN 2017 WSCC32
  "Zooming into dark matter and proto-galaxies with massive lensing
  clusters".  MG acknowledges financial support from the grant MIUR
  PRIN 2022 KCS97B: ``EMC2 Euclid Mission Cluster Cosmology:
  unlock the full cosmological utility of the Euclid photometric
  cluster catalog” and financial support from the University of
  Trieste through the programs FRA~2023 and FRA~2024. AB acknowledges financial
  support from the INAF mini-grant 1.05.12.04.01 ``The dynamics of
  clusters of galaxies from the projected phase-space distribution of
  cluster galaxies''. RD gratefully acknowledges support by the ANID BASAL project FB210003.

  \\ This publication is based on observations collected at the
  European Organisation for Astronomical Research in the Southern
  Hemisphere under ESO programme IDs 186.A-0798 and 096.A-0105.
  
  \\ This publication is partly based on observations made on the island
  of La Palma with the Italian Telescopio Nazionale Galileo (TNG),
  which is operated by the Fundaci\'on Galileo Galilei -- INAF
  (Istituto Nazionale di Astrofisica) and is located in the Spanish
  Observatorio of the Roque de Los Muchachos of the Instituto de
  Astrof\'isica de Canarias.

  \\The DESI Legacy Imaging Surveys consist of three individual and
  complementary projects: the Dark Energy Camera Legacy Survey
  (DECaLS), the Beijing-Arizona Sky Survey (BASS), and the Mayall
  z-band Legacy Survey (MzLS).  The
  complete acknowledgments can be found at
  https://www.legacysurvey.org/acknowledgment/.\\
  The Photometric Redshifts for the Legacy Surveys (PRLS) catalog used
  in this paper was produced thanks to funding from the
  U.S. Department of Energy Office of Science, Office of High Energy
  Physics via grant DE-SC0007914.

\end{acknowledgements}
    
\bibliographystyle{aa}
\bibliography{biblio}

\end{document}